\documentclass[aps,prd,showpacs,superscriptaddress,groupedaddress,nofootinbib,amsmath]{revtex4}  

\usepackage{graphicx}  
\usepackage{dcolumn}   
\usepackage{bm}        
\usepackage{amssymb}   
\usepackage{amsbsy}
\usepackage{mathtools}
\usepackage{color}
\usepackage{wasysym}
\usepackage{eso-pic}
\usepackage{graphicx}
\usepackage{color}
\usepackage{type1cm}

\begin{document}

\title{An exact result concerning the $1/f$ noise contribution \\to the large-angle error 
in CMB temperature and polarization maps}

\author{Martin Bucher}
\email{bucher@apc.univ-paris7.fr} 

\affiliation{Laboratoire APC, Universit\'e Paris 7/CNRS, B\^atiment Condorcet, 
Case 7020, 75205 Paris Cedex 13, France}

\affiliation{Astrophysics and Cosmology Research Unit,
School of Mathematics, Statistics and Computer Science \\
University of KwaZulu-Natal
Durban 4041, South Africa}


\def\gtorder{\mathrel{\raise.3ex\hbox{$>$}\mkern-14mu
             \lower0.6ex\hbox{$\sim$}}}
\def\ltorder{\mathrel{\raise.3ex\hbox{$<$}\mkern-14mu
             \lower0.6ex\hbox{$\sim$}}}

\date{\today}

\begin{abstract}
\noindent
We present an exact expression for the $1/f$ contribution to the noise of the CMB temperature 
and polarization 
maps for a survey in which the scan pattern is isotropic. The result for polarization 
applies likewise to 
surveys with and without a rotating half-wave plate. A representative range 
of survey parameters is explored and implications for the design and optimization of future 
surveys are discussed. These results are most directly applicable to space-based surveys, which 
afford considerable freedom in the choice of the scan pattern on the celestial sphere.
We discuss the applicability of the methods developed here to analyzing past experiments and
present some conclusions pertinent to the design of future experiments. 
The techniques developed here 
do not require that the excess low frequency noise have exactly the 
$1/f$ shape and readily generalize to other functional forms for 
the detector noise power spectrum. 
In the case of weakly anisotropic scanning patterns the techniques in this paper
can be used to find a 
preconditioner for solving the map making equation 
efficiently 
using the conjugate gradient method. 
\end{abstract}

\maketitle

\section{Introduction}

Most simple forecasts for future CMB experiments postulate a white isotropic instrument noise, 
a hypothesis under which extremely simple and almost trivially derived expressions for the 
uncertainties and the form of the likelihood function follow \cite{knox,martinReview}. However this 
model often greatly underestimates the uncertainties in the measured anisotropies on large 
angular scales, for which the contribution of $1/f$ noise often dominates over the white noise 
contribution, which arises from both the intrinsic fluctuations of the incoming photons and the 
uncorrelated white noise added by the detector.
For studies of the CMB temperature anisotropies, because of the red shape of the primordial CMB 
spectrum, with $C_{TT}\sim \ell ^{-2}$ (ignoring the acoustic peaks, damping tails, and all 
that), the system requirements for measuring the large scales are less critical than for surveys 
targeting the polarized signal. With detectors sufficiently sensitive to measure the primordial 
temperature spectrum on the smallest angular scales accessible to the instrument, 
the sensitivity in temperature on large 
scales comes almost for free, and quite a significant increase in the noise on large 
angular scales above the 
level that would result 
from the white noise alone can be tolerated. However the shape of the polarization power 
spectrum for $\ell \ltorder 100$ is roughly $c_{EE}, c_{BB}\sim \ell ^0$ (again ignoring 
subtleties such as the reionization bump and the fall off at larger $\ell $). Its 
shape at low $\ell $ resembles a white noise power spectrum. This implies that any increase 
in the $1/f$ noise relative to the level of the 
white noise can compromise polarization measurements on large angular scales. Consequently,
accurately estimating the errors on these scales is of the utmost importance. Indeed 
the lowest multipoles of the $c_{BB}$ power spectrum is where the so-called `reionization bump' 
is located. This is one of the most sensitive and promising windows for detecting the primordial 
tensors or primordial gravitational waves that presumably were generated during the epoch of cosmic inflation 
\cite{baumann,coreWhitePaper,reionBump}.

Although this fact is frequently obscured by the prevalence of analyses using a white noise model 
for the measurement error in CMB surveys, all CMB measurements to date in one way or another 
have relied on differential measurements of the sky temperature as measured in a given 
frequency band to produce maps of the CMB anisotropy. The idea of always exploiting 
differential measurements dates back to Robert Dicke \cite{dicke}. The COBE DMR experiment 
\cite{cobe}, for example, used two beams defined by horns pointing in directions separated by 
$60^\circ $ on the sky. The detector was electronically rapidly switched between the two 
horns, and only the data consisting of differences between the detectors was retained in order 
to reject the wandering of the detector zero point. With improvements in technology, the 
timescale after which such drifts start to contribute substantially to the noise has greatly 
increased, thus enabling surveys to difference less directly by relying on 
the motion of the beam through the sky
rather than on hardware switching.
Most modern CMB experiments rely on such differencing, which is implemented implicitly 
in the analysis rather than directly in the hardware, as we shall now explain. For example, the 
Planck HFI (High Frequency Instrument) \cite{planckHFI} carried out continuous scans across the 
sky with no switching. Nevertheless, because of the presence of such zero point drifts, also 
known as $1/f$ noise, the sky maps constructed from the Planck survey and from surveys with 
data from other experiments are also based on combining differential measurements.

Most treatments including $1/f$ noise rely on complicated simulations that are highly demanding 
in computational resources because matrix equations of high dimension [i.e., 
O(${n_{pixel}}$)] must be solved. Here, however, we show how for isotropic scan patterns, simple 
results may be obtained without resorting to any linear algebra, but rather by exploiting the 
symmetry of the scan pattern. While simulations taking into account the full complexity of a 
real experiment are unavoidable, we hope that the methods presented here may provide invaluable 
intuition and estimates of what to expect from more comprehensive simulations and from results 
for anisotropic scan patterns.

\section{Map making and $1/f$ noise: the temperature case}

A CMB detector with zero point drifts may be idealized by means of 
a time series of measurements
subject to errors described as a stationary Gaussian white noise, which is 
completely determined through its power spectrum.
Formally we may express the data vector $\mathbf{d}$ through the equation 
\begin{equation}
\mathbf{d}
=
\mathbf{A}~
\mathbf{m}
+
\mathbf{n}
\end{equation}
where $\mathbf{m}$
represents the pixelized sky map,
$\mathbf{A}$ the pointing matrix, and 
$\mathbf{n}$ the noise vector.
According to the map making equation (see for example, \cite{mapMaking}), 
the maximum likelihood sky map is given by 
\begin{equation}
\mathbf{m}_{ML}=
(\mathbf{A}^T \mathbf{N}^{-1} \mathbf{A})^{-1}
(\mathbf{A}^T \mathbf{N}^{-1}) \mathbf{d}
\label{mapMakingEquation}
\end{equation}
where $\mathbf{N}$ is the noise covariance matrix of the detector.\footnote{Here we retain
an abstract notation and do not specify the exact content of the 
vectors $\mathbf{m}$ and $\mathbf{d}.$ 
For polarization insensitive measurements, the vector $\mathbf{m}$ might consist of 
a pixelized temperature sky map and the data vector $\mathbf{d}$ might consist of the detector
output sampled at a certain finite rate. On the other hand, for a partially linearly polarized
sky measured with polarization sensitive bolometers, 
the vector $\mathbf{m}$ would instead consist of a three sky maps, one for each of the three
relevant Stokes parameters $I,$ $Q,$ and $U,$ 
and the data vector $\mathbf{d}$ would consist of the detector output measuring a particular
linear polarization in a certain direction on the sky. In this case the pointing matrix would
also encode the instantaneous orientation of the polarizer on the sky.}
Here we shall
ignore issues of discretization or pixelization, which in practice are 
very important but need not be considered for obtaining the exact
results reported here. If we regard the detector output
as a continuous time stream, we may characterize its noise by its frequency
power spectrum, defined as 
\begin{equation}
\langle n(\nu )~n(\nu ')
\rangle =N(\nu ) \delta (\nu -\nu '),
\end{equation}
which we 
take to have the form
\begin{equation}
N(\nu )=N_{white}\left( 1+\frac{\nu _{knee}}{\nu }\right) .
\label{noiseWhitePlusOneOverF}
\end{equation}
In other words, we assume a white noise component with a $1/f$ noise component superimposed.
It is convenient to parameterize the amplitude of the $1/f$ component through 
its `knee frequency' $\nu _{knee},$ defined as the frequency at
which the $1/f$ noise starts to dominate over the white 
noise. (For a discussion of the measured noise spectrum in CMB detectors, see for example
\cite{cmbDetectorNoise}.)

Experimentally, it has been observed that the noise at low frequencies has a power spectrum well
approximated by an inverse power law $1/f^\alpha $ where $\alpha $ is close to one. Such noise 
has been observed in a wide variety of measurement contexts, not just in CMB 
detectors \cite{oneOverF}. This so-called $1/f$ noise seems to be universal, although its 
physical origin is not well understood and lacks a convincing, generally applicable theoretical 
explanation.  We note that many of the results and techniques developed here can readily be 
generalized to the case where $\alpha $ is not exactly one. However for concreteness 
we do not explore this generalization here, instead restricting ourselves to the case $\alpha =1.$

\begin{figure}
\includegraphics[width=3.5in]{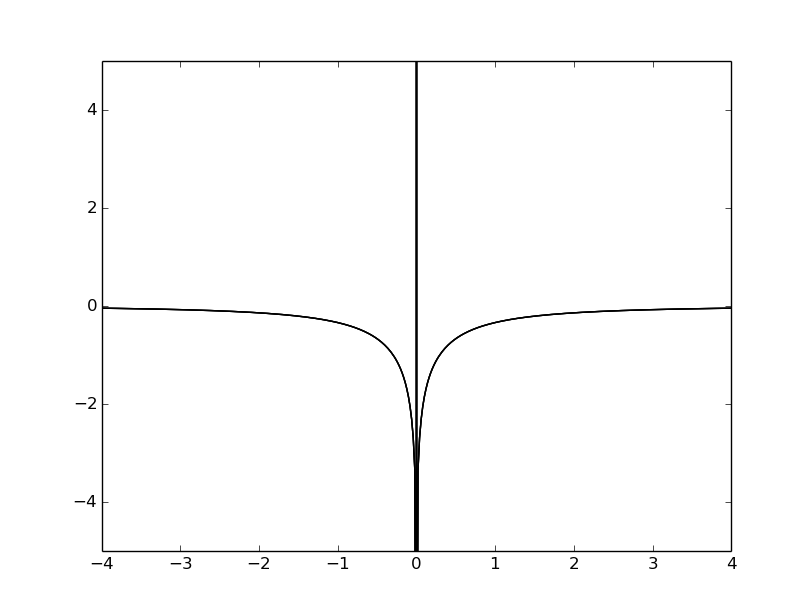}
\caption{\label{nMinusOnePlot} 
{\bf Shape of the high-pass map making filter.}
In the process of optimal map making as carried out using the map making
equation [eqn.~(\ref{mapMakingEquation})], the time ordered data stream is convolved with
the high pass filter $N^{-1}(t),$ as explained in the main text.
Under the assumption of combined white and $1/f$ noise, the shape of this filter has a universal form 
having only one undetermined parameter, the filter width, which is given by ${\nu _{knee}}^{-1}.$
The spike in the center represents a unit delta function and the total area of the negative region 
excluding the singularity at the origin is $-1,$ thus providing a perfect cancellation. 
The horizontal axis is $\tau =2\pi \nu _{knee}t$ and the vertical axis is dimensionless.
}
\end{figure}

From the form of the noise spectrum in eqn.~(\ref{noiseWhitePlusOneOverF}), it is apparent that 
the temperature measurements taken are not absolute in character. Mathematically, if we 
regularize the high frequency divergences inherent in white noise 
by integrating over a pixel of a finite width $(\Delta t),$ we 
are left with logarithmic divergences of the form $\int d\nu /\nu $ at low frequencies. 
Physically, this divergence reveals a complete uncertainty regarding the value of the absolute 
zero point. Differences between successive measurements, however, do not suffer from this 
infrared divergence, and for large $t$ the expectation value of the difference squared grows in 
proportion to $\log (t)$ \cite{janssen} where $t$ is the time between the two measurements.

We could try to regularize this divergence, but there is no need to
do so because it is $N^{-1}(t)$ and not $N(t)$ that enters into
the map making equation. 
We may integrate
\begin{eqnarray}
&&N^{-1}(t)=2\int _0^\infty d\nu ~N^{-1}(\nu )~\cos (2\pi \nu t)
=2{N_{white}}^{-1}
\int _0^\infty d\nu ~\left( \frac{\nu }{\nu +\nu _{knee}}\right) ~\cos (2\pi \nu t),
\end{eqnarray}
obtaining an analytic result in terms of special functions, namely functions
related to the sine and cosine integral, so that
\begin{equation}
N^{-1}(t)={N_{white}}^{-1}
\Bigl[ \delta (t)- 2\nu _{knee}~g\bigl( 2\pi \nu _{knee}\vert t\vert \bigr) \Bigr] ,
\end{equation}
which is plotted in Fig.~\ref{nMinusOnePlot}.
Here $g(x)$ may be defined by the integral
\begin{equation}
g(x)=\int _0^\infty \frac{dw~\cos (w)}{(w+x)}.
\end{equation}
The special function 
$g(x)$ is known as one of the two auxiliary functions of the sine and cosine integrals
[see chapter 5 of  Abramowitz \& Stegun \cite{abramowitz}, in particular eqn.~(5.2.13), 
or the Digital Library of Mathematical Functions \cite{dlmf} (hereafter DLMF), 
in particular eqn.~(6.7.14)]. This function can also be expressed in terms of the 
sine and cosine integrals according to [DLMF, eqn.~6.2.18]
\begin{equation}
g(z)=
-\textrm{Ci}(z)\cos (z)
-\textrm{si}(z)\sin (z).
\label{gDefinition}
\end{equation}
We note that\footnote{This 
follows from the identities [see DLMF, 6.14.5]:
\begin{equation*}
\int _0^\infty dw~\sin (w)~\textrm{si}(w)
=\int _0^\infty dw~\cos (w)~\textrm{Ci}(w)
=-\frac{\pi }{4}.
\end{equation*}
}
$\int _0^\infty dx~g(x)={\pi }/2,$
so that 
\begin{equation}
\int _{-\infty }^{+\infty }dt~N^{-1}(t)=0.
\label{EqEight}
\end{equation}
For small $x,$ $g(x)\approx -\ln (x)+\gamma $ where $\gamma =0.577$ is the Euler-Mascheroni
constant, and for large $x,$ $g(x)\approx 1/x^2.$
Although this function diverges at the origin, its singularity is mild because it is only
logarithmic. Consequently, when integrated over, this singularity disappears and thus
does not pose any problem for us below. We have no need to regularize it.

We may consider $N^{-1}(t)$ as it appears in the map making equation as 
a high pass filter, allowing components in the data time stream with $\nu \gg \nu _{knee}$
to pass almost without attenuation or  distortion, but almost completely blocking the low frequency 
components with $\nu \ll \nu _{knee}$ 
and completely filtering out the $\nu =0$ component.  Map making
consists of two steps: first an intermediate map is populated with the data, distributed
along the scans according to $N^{-1}(t),$ so that 
\begin{equation}
\mathbf{m}_{int}=\mathbf{A}^T
\mathbf{N}^{-1}\mathbf{d},
\end{equation}
and then this intermediate map is corrected and renormalized according to 
\begin{equation}
\mathbf{m}_{ML}=(\mathbf{A}^T\mathbf{N}^{-1}\mathbf{A})^{-1}\mathbf{m}_{int}.
\phantom{\int }
\label{mapMakingEquationBis}
\end{equation}
Because of the property in eqn.~(\ref{EqEight}), the average of the intermediate map is zero.
Moreover a constant map is annihilated by the operator $(\mathbf{A}^T\mathbf{N}^{-1}\mathbf{A}),$ 
so for the map making equation to be well posed,
the dimension spanned by the constant map must be removed from the linear space
of possible maps.
The matrix $(\mathbf{A}^T\mathbf{N}^{-1}\mathbf{A})$ is the inverse noise or information matrix 
for the map resulting from the above procedure, and below we shall focus on 
calculating the properties of the noise in the final reconstructed map that is
encoded in this matrix.
For the most general scan pattern, neither the form of the intermediate map, nor that of the 
solution of the map making equation is particularly intuitive. In this general case, the map 
making equation almost magically yields the best possible map given the survey data.

There is, however, a special class of surveys for which a simple characterization of the 
intermediate map and of the final noise is possible: the class of isotropic scan surveys. We 
define an {\it isotropic scan survey} 
as follows. Formally, a survey is isotropic if the distribution 
of scans through a pixel is invariant under rotations of the pixel and also under rotations 
of the celestial sphere 
interchanging pixels. Here pixel could mean a point on the celestial sphere or the pixel center 
in a commonly used pixelization scheme such as {\tt Healpix} \cite{healpix}. 
Strictly speaking, the scans of a 
survey can satisfy the requirement of isotropy only in a sort of formal continuum limit, which 
would require an infinite number of scans. But we shall ignore these complications, which are 
not especially relevant for our purposes, which is understanding the impact of $1/f$ noise on 
the largest angular scales of the map. For our purposes, we are able to obtain results working in a 
continuum limit, without regard to a particular pixelization.

A particular example of an isotropic survey will be used as a worked example for which numerical
results will be given. 
This worked example assumes a survey where scans 
are taken along circles of opening angle (or spherical radius) $\beta .$ If $\Omega $ is the angular 
speed of the beam on the sky, we may define a `knee angle' 
$\Theta _{knee}=\Omega \nu ^{-1}_{knee}.$ We assume that the scan circles are uniformly distributed on 
the celestial sphere, that each circle is traversed many times (to avoid 
complications at the endpoints), and that the total circumference traversed on each circle is 
much larger than $\Theta _{knee}.$ Under these assumptions, for this family of surveys, only 
the two parameters $\beta $ and $\Theta _{knee}$ are relevant to determining the $1/f$ 
enhancement of the final sky map noise.

This worked example is
in several respects not so different from the scan 
pattern described in the EPIC proposal \cite{epic,epicBis}, where the beam sweeps around 
circles in the sky around the spin direction of the satellite, or the `boresight' direction, 
which in turn undergoes a precession at a much slower rate. (See also the descriptions of the 
contemporaneous SAMPAN \cite{sampan} and B-Pol \cite{bpol} proposals where a similar scanning 
pattern of this sort was also proposed.) In the EPIC proposal, the circles do not close because of 
the precession, but one can also consider a similar scan pattern where the circles close 
exactly, with the precession occurring in discrete steps.

For an isotropic scanning pattern, the final map inverse noise 
operator 
$\mathbf{A}^T\mathbf{N}^{-1}\mathbf{A}$
is isotropic and as a consequence takes the particularly simple form
\begin{eqnarray}
&&
\Bigl( \mathbf{A}^T\mathbf{N}^{-1}\mathbf{A}\Bigr) (\hat \Omega , \hat \Omega ')
\phantom{\int }
=({\cal N}_{white})^{-1}
\sum _{\ell =0}^\infty  
w_{\ell }^T 
\sum _{m=-\ell }^{+\ell } 
Y_{\ell m}(\hat \Omega )
Y^*_{\ell m}(\hat \Omega '),\quad 
\label{oneEq}
\end{eqnarray}
where $w_\ell ^T\to 1$ as $\ell \to \infty .$ 
For small $\ell ,$ $w_\ell <1,$ 
and $w_0=0.$
The form on the right-hand side of eqn.~(\ref{oneEq})
arises because the eigenfunctions of an isotropic 
operator are simply the spherical harmonics
$Y_{\ell m}(\hat \Omega ),$ and the corresponding eigenvalues can
depend only on $\ell $ as a consequence of the isotropy.
The sum over $m$ on the right is a projection
operator onto the linear subspace of angular momentum $\ell .$
The coefficients $w_\ell ^T$ are given by 
\begin{eqnarray}
w_\ell ^T&=& 1 - \sum _{(i)=1}^nf_{(i)}\int _{-\infty }^{+\infty }dt~(2\nu _{knee})~
g(2\pi \nu _{knee}\vert t\vert )~
P_\ell \Bigl( \cos [\theta _{(i)}(t)]\Bigr)
\label{wTemp}
\end{eqnarray}
where $P_\ell(\cdot )$ is the Legendre polynomial of order $\ell .$ 
In the above expression we have assumed 
that there are $n$ types of scans
passing through a reference pixel (here taken to be situated at the north pole), 
the scan of the type $(i)$ 
occurring a fraction $f_{(i)}$ of the time,
so that $\sum _{(i)=1}^{n}f_{(i)}=1.$
Here $t$ is the time taken relative to the moment of transit through the reference pixel,
and $\theta _i(t)$ is the angle of the pixel visited at time $t$ relative to
the reference pixel. Of course, because of the isotropy the choice of reference pixel does not matter.
${\cal N}_{white}$ is the limiting white noise variance of the 
sky map on small angular scales, where $1/f$ noise is irrelevant.\footnote{
The overall normalization ${\cal N}_{white}$ of the
noise power spectrum of the final map 
is related to individual detector white noise amplitude
$N_{white},$ which has units of
$(\textrm{Temperature})^2\cdot (\textrm{Time}),$ commonly
expressed in units of $\mu K^2\cdot s.$
The relation is
\begin{equation*}
{\cal N}_{white}=(4\pi )^{1/2}
{N_{white}}/({n_{det}~t_{survey} })
\end{equation*}
where
$n_{det}$ is the total number of detectors
and the time
$t_{survey}$ is the duration of the survey. The $(4\pi )^{1/2}$
factor arises from
the normalization convention of
the spherical harmonics.
If the detectors had only white noise with the superimposed low-frequency $1/f$ noise component 
turned off, the spherical harmonic expansion coefficients of the noise in the final map
would have the two-point correlations
\begin{equation*}
\left<  n_{\ell m}^{\phantom{*}}~ n_{\ell ' m'}^* \right>
={\cal N}_{white}~\delta _{\ell \ell '}~\delta _{mm'}.
\end{equation*}
}

The correctness of eqn.~(\ref{wTemp}) may be demonstrated 
in the following way.
For the sector with angular momentum $\ell ,$ we can solve for $w_\ell $ by means of the 
eigenvalue equation
\begin{equation}
\int d\hat \Omega '~
\Bigl( \mathbf{A}^T \mathbf{N}^{-1} \mathbf{A}\Bigr) (\hat \Omega , \hat \Omega ')~P_\ell (\cos \theta ')
=
{{\cal N}_{white}}^{-1}~
w_\ell ~P_\ell (\cos \theta ).
\label{evEqnTemp}
\end{equation}
Here we have used the fact that 
$Y_{\ell 0}(\theta , \phi )=
\sqrt{(2\ell +1)/(4\pi )}~P_\ell (\cos \theta )$ is an eigenvalue of the 
operator. 
Setting $\theta =0,$ we may rewrite
\begin{eqnarray}
{{\cal N}_{white}}^{-1}~
w_\ell ~
&=&
\int d\hat \Omega ~
\int _{t_{start}}^{t_{end}}dt~
\delta ^2\bigl( \Omega (t), \Omega _{N.P.}\bigl) 
\int _{-\infty }^{+\infty }dt'~N^{-1}(t-t')~
P_\ell \bigl( \cos [\theta (t')]\bigr) \cr 
&=&
{{\cal N}_{white}}^{-1}~
\sum _{(i)}f_{(i)}~
\int _{-\infty }^{+\infty }dt'~h(t')~
P_\ell \bigl( \cos [\theta _{(i)}(t')]\bigr) .
\label{evEqnTempB}
\end{eqnarray}
Here $\delta ^{(2)}(~\cdot ~, ~\cdot ~)$ 
is the two-dimensional $\delta $-function, normalized with respect to the usual
area element on the unit sphere---in other words, 
$\delta ^{(2)}(\hat \Omega , \hat \Omega ')=
(1/\sin \theta )~\delta (\theta -\theta ')~\delta (\phi -\phi ').$ 
In the above we ignore endpoint effects, whose fractional contribution would be suppressed by a factor 
of $\nu _{knee}(t_{end}-t_{start}).$
In the bottom line we have shifted the origin of the time coordinate $t'$ so that $t'=0$ coincides with
the crossing at the north pole $\theta =0,$ and we also assume (in using the infinite range of integration)
that the various crossings do not overlap---in other words, $h(t)$ has decayed sufficiently so that
there is no overcounting.
Here $h(t)$ is normalized so that its singular part at the origin is equal to $\delta (t).$
For the particular case analyzed in this paper where
$h(t)=(2\nu _{knee})~g(2\pi \nu _{knee}\vert t\vert ),$ we may rewrite the above as 
\begin{eqnarray}
(1-w_\ell )&=&
\sum _{(i)}f_{(i)}~
\int _{-\infty }^{+\infty }dt'~
(2\nu _{knee})~g(2\pi \nu _{knee}\vert t\vert )~
P_\ell \bigl( \cos [\theta _{(i)}(t')]\bigr) .
\end{eqnarray}
The extension to a more general form for the low-frequency noise (or slightly colored high-frequency noise)
is straightforward. 

Here we have used a continuum notation, without resorting to any particular pixelization. But the step
\begin{eqnarray}
\int _{t_{start}}^{t_{end}}dt ~
\int _{\Omega _{pix},N.P.}d\hat \Omega ~
\int _{-\infty }^{+\infty }dt~
\delta ^2\bigl( \Omega (t), \Omega _{N.P.}\bigl) 
\ldots
&=&
T_{pix}~
\sum _{(i)}f_{(i)}~
\ldots 
\end{eqnarray}
may be analyzed in the following way using a pixelization with a pixel of area $\Omega _{pix}$ centered about
the north pole (i.e., $\theta =0$). The integral on the left-hand side expresses the total time 
$T_{pix}$ spent by the beam in this pixel $\Omega _{pix}$ whose area is denoted by the same symbol.
The sum on the right-hand side divides the time spent in this pixel into classes, here formally labelled
by the discrete index $(i),$ but for certain scanning patterns a continuous index as well would be required.
For scanning using closed circles, as assumed in the worked example, one would in principle have to separate
the scans crossing the reference pixel according to their direction of transit, but in the above expression 
where $P_\ell (\cos \theta )$ is azimuthally symmetric, the crossing angle can be ignored. For 
a single continuous scan where the center of the spin axis precesses or otherwise moves, 
the radius of curvature would
vary, and in principle this subclassification would be required to obtain an exact expression. But
when the precession rate is small compared to the spin rate, it is a good approximation to ignore this
complication.

Isotropy is crucial to obtaining the above simple form for the noise in the sky map. 
In the absence of isotropy, the 
operator in eqn.~(\ref{oneEq}) acquires nonzero off-diagonal elements, rendering the calculation 
substantially more complex, and for the non-isotropic case
substantial computing resources are needed to deal with the 
linear algebra, which is of extremely high dimension.

We note that even though the worked example described above involves separate closed circles, 
the result in eqn.~(\ref{wTemp}) also can be applied to scan patterns where the circles do not 
close---that is, for a scan pattern with a spin and a slower precession 
of the spin axis superimposed, so that 
the center of the circle wanders in a continuous manner. 
The only requirement, at least for obtaining a result that is 
exact, is that this precession, or wandering, is itself isotropic. We do not explicitly analyze 
such scan patterns here, because in the limit where the precession is very slow, the results 
become identical to the worked example with closed circles.

In the above analysis, we do not take into account the nonzero beam width,
which acts to smear or attenuate the CMB sky signal on angular scales small compared to the 
beam width. The noise is superimposed additively onto a sky map representing
the actual sky as seen through a low-pass filter whose properties are 
determined by the precise 
beam profile. However the properties of this superimposed noise 
do not depend on the details of the beam and therefore do not have to be 
considered here. The only crucial simplifying assumption is the azimuthal symmetry of the beam 
about its center.

\section{Numerical results: temperature case}
\label{SectNumTemp}

\begin{figure}
\vskip -0.15in 
\begin{center}
\includegraphics[width=0.5\textwidth]{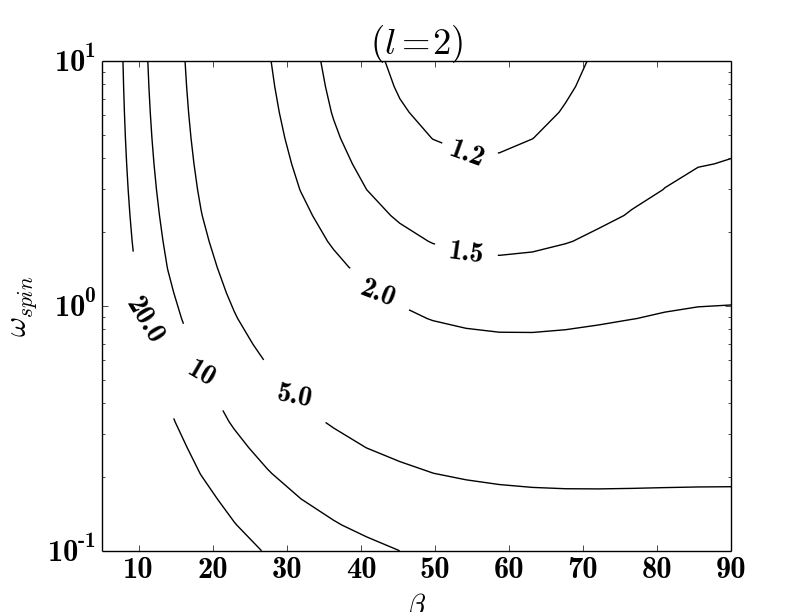}\\
\includegraphics[width=0.5\textwidth]{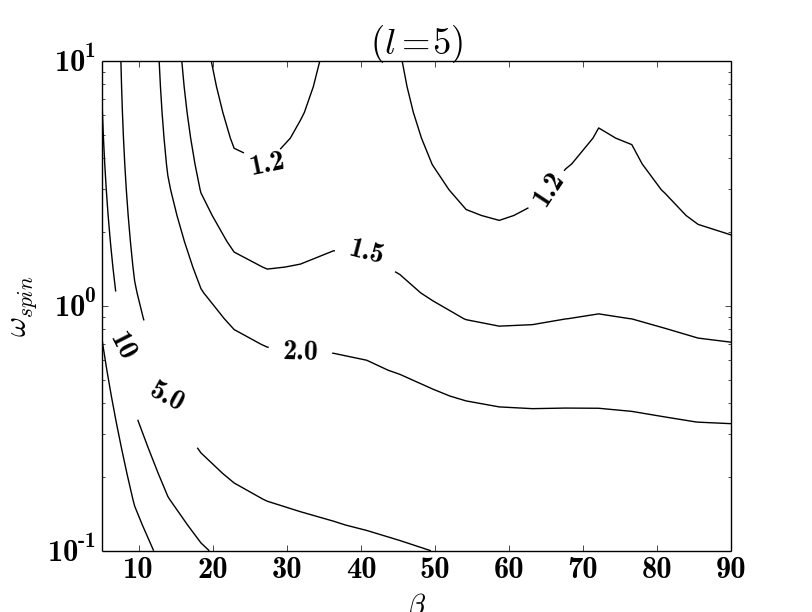}\\ 
\includegraphics[width=0.5\textwidth]{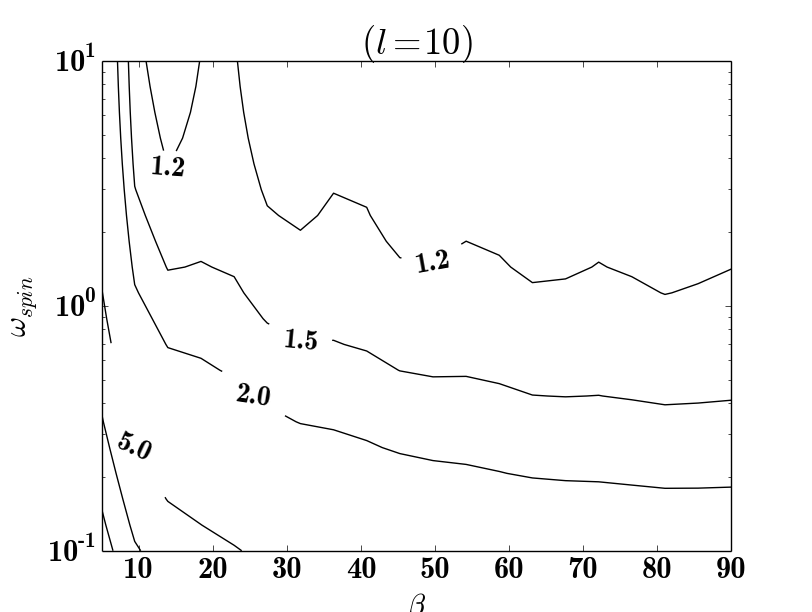}\\
\end{center}
\vskip -0.15in 
\caption{\label{numFig} {\bf Temperature $1/f$ noise boost factor.} 
Using a logarithmic scale, we show the multiplicative factor $B_\ell ={(w_\ell ^T)}^{-1}$ by which the 
total noise is boosted relative to what the noise would be if there were only the white
noise, according to the expression in eqn.~(\ref{workedExampleTemp}).
Above the boost factor $B_\ell $ is shown (from top to bottom) 
for $\ell =2, 5,$ and $10.$
Here $\beta $ is the radius (expressed in degrees) 
of the scanning circles and $\omega _{spin}=\Omega _{spin}/(2\pi \nu _{knee})$
is the scanning rate compared to the $1/f$ knee frequency.
}
\end{figure}

Above we described a worked example where scans consisted of circles traversed many times whose
centers are distributed uniformly on the celestial sphere. In this Section, using the formulas
derived in the previous section, we present explicit numerical results for this worked example.
To proceed, we require the explicit trajectory of a scan with its 
the reference point situated at the `north pole,' or at $O=(0, 0, 1),$ which is as
follows:
\begin{equation}
\begin{pmatrix}
\hat n_x(t)\\ 
\hat n_y(t)\\ 
\hat n_z(t)
\end{pmatrix}
=
\begin{pmatrix*}[l]
1 ~~  & 0 & 0\\
0 ~~  & \cos \beta  & -\sin \beta \\
0 ~~  & \sin \beta  & \phantom{-}\cos \beta
\end{pmatrix*}
\begin{pmatrix}
\sin \beta  ~\sin (\Omega _{spin}t)\\
\sin \beta  ~\cos (\Omega _{spin}t)\\
\cos \beta  
\phantom{~\sin (\Omega _{spin}t)}
\end{pmatrix}
=
\begin{pmatrix*}[l]
\sin \beta ~\sin (\Omega _{spin}t)\\
\sin \beta ~\cos \beta \bigl( \cos (\Omega _{spin}t)-1\bigr) \\
\sin ^2 \beta \cos (\Omega _{spin}t)+\cos ^2\beta 
\end{pmatrix*}.
\label{spinPattern}
\end{equation}
The component $\hat n_z(t)=
\cos ^2\beta + \sin ^2\beta \cos (\Omega _{spin}t)$
corresponds to the $z=\cos \theta $ that appears
as the argument of the Legendre polynomial in eqn.~(\ref{wTemp})
and, further below, of the polar mode functions ${\cal P}_j(z)$ 
and ${\cal Q}_j(z)$ in eqn.~(\ref{wPol}) for the case of polarization.
Eqn.~(\ref{spinPattern}) may also be expressed in the more compact form
\begin{equation}
\hat {\mathbf{n}}(\tau )=
\boldsymbol{\mathcal{R}}_{\hat {\mathbf{x}}}(-\beta )~
\boldsymbol{\mathcal{R}}_{\hat {\mathbf{z}}}(\Omega _{spin}t)~
\boldsymbol{\mathcal{R}}_{\hat {\mathbf{x}}}(+\beta )~
\hat {\mathbf{e}}_z,
\label{noprecEq}
\end{equation}
where $\boldsymbol{\mathcal{R}}_{\hat {\mathbf n}}(\theta ) $
is a rotation by an angle $\theta $ about the axis $\hat {\mathbf n}.$

For the class of idealized scans described above, the results for the $1/f$ noise variance 
boost factor $B_\ell ^T=(w_\ell ^T)^{-1}$ depends only on two parameters: the opening angle $\beta $ 
of the scan circles and dimensionless scanning speed $\omega ={\Omega _{spin}}/({2\pi \nu _{knee}})$ 
where the $1/f$ knee frequency is used as a reference for comparison.
For the scan pattern for the worked example described above having  
the trajectory in eqn.~(\ref{spinPattern}), 
eqn.~(\ref{wTemp}) becomes 
\begin{eqnarray}
&&\Bigl( 1-w_\ell ^T\Bigr) =
\frac{1}{\pi }
\int _{-\infty }^{+\infty }dT~g(\vert T\vert )~
P_\ell \left( 
\cos ^2\beta + \sin ^2\beta \cos \left( \omega T\right)
\right) .
\label{workedExampleTemp}
\end{eqnarray}
Here 
$\omega $ is defined as indicated above 
after the rescaling $T=2\pi \nu _{knee}t.$ 
In Fig.~\ref{numFig} we show the results 
for $(w^T_2)^{-1},$ $(w_5^T)^{-1},$ and $(w^T_{10})^{-1}$ as a function of these two parameters.

For the precessing case the radius of curvature 
varies periodically along the scan. Therefore, to obtain an exact result, it is necessary to average
over position along the scan.
In the limit where the precession is slow compared to the spin,
one would recover the result in eqn.~(\ref{workedExampleTemp}). 
Unless the precession is faster than or of the same order as the spin, the result for the precessing case
will not differ significantly from the non-precessing case because whether or not the circles close
does not enter in an essential way into the calculation. There may however be strong arguments based
on considerations beyond the scope of this paper, both in favor of closed circles and in favor of circles 
that do not close because of a precession of the spin axis. We note that closed circles provide 
a wealth of exactly redundant null tests, thus providing invaluable checks of systematic errors and a characterization
of the noise at different epochs during the survey. However, for a given survey length, this redundant information comes 
at the expense of less dense coverage on the celestial sphere, which could result in the aliasing of small scale
power into large scale power. Moreover, in addition to some dead time required to reposition the 
spin axis, there are also endpoint effects at the beginning and end of each closed circle scan during which
the offset is no longer as well constrained by the data as it would otherwise be. These endpoint transients
last a time of order $t\approx (\nu _{knee})^{-1}.$

\section{Generalization to polarization}

For concreteness, we first specify the simple setup assumed here 
for measuring the temperature and 
polarization to which the analysis below applies. 
We have a single detector sensitive to only one linear 
polarization, so that without noise the detector would measure
\begin{equation}
T(t)=I(t)+\cos 2\xi (t)~Q(t) +\sin 2\xi (t)~U(t),
\end{equation}
where $\xi (t)$ indicates the orientation of the 
linear polarization measured by the detector and $I(t),$ $Q(t),$ 
and $U(t)$ are the Stokes parameters of the sky integrated over the profile of 
the beam on the sky at time $t.$
We idealize an azimuthally symmetric beam and 
a perfect detector with 
absolutely no leakage from the other polarization.
The analysis can also be straightforwardly generalized to variations of 
this basic setup.

The symmetry of an isotropic scan pattern can be exploited to obtain an expression 
analogous to eqn.~(\ref{wTemp}) for the cases both with and without a rotating half-wave plate 
(HWP). 
For the polarized case, in order to simplify the analysis, 
we strengthen the hypothesis of isotropy to require that
the scan pattern also be nonchiral---that is, invariant not
just under proper rotations but also under reflections
about great circles, which for a sphere embedded in
three-dimensional Euclidean space would correspond
to reflections about planes passing through the origin.
Because of the assumed symmetry, the operator
\begin{eqnarray}
{\cal O}=(\mathbf{A}^T\mathbf{N}^{-1}\mathbf{A})
\end{eqnarray}
that arises 
for the polarized case also has total angular momentum zero.
Therefore this operator
acting on a function of definite angular momentum defined by
the quantum numbers $jm$ conserves these quantum numbers.

The most general operator invariant under proper rotations, however,
can mix the modes 
$E_{jm}$ and $B_{jm}.$ To see how this is possible, we consider
the isotropic operator that rotates the second-rank tensor on
the sphere by an angle $\zeta $ about $\hat {\mathbf{e}}_r.$
Its action on the polarization eigenfunctions of total angular momentum $j$
is given by 
\begin{equation}
\begin{pmatrix}
E_{jm}\\
B_{jm}
\end{pmatrix}
\to
\begin{pmatrix}
\phantom{-}\cos 2\zeta & \phantom{+}\sin 2\zeta \\
-\sin 2\zeta &
\phantom{-}\cos 2\zeta
\end{pmatrix}
\begin{pmatrix}
E_{jm}\\
B_{jm}
\end{pmatrix}.
\end{equation}
More generally, a different angle $\zeta _j$ may occur for
each $j.$ To forbid such mixing, more symmetry is required.
We can forbid such mixing by also requiring invariance under 
improper rotations because $E_{jm}$
and $B_{jm}$ have opposite parities. An S-shaped scan,
for example, would violate parity by selecting a preferred chirality
and thus would generically mix $E$ and $B$ modes of the same $jm$
quantum numbers. 

Let us for the moment (assuming the absence of circular or elliptic polarization) 
assume a partially linearly polarized sky described by 
the Stokes parameters $I,$ $Q,$ and $U$ in some suitable basis. 
Because of the isotropy of the scan pattern, the mode functions 
$T_{jm,ab}(\hat \Omega ),$
$E_{jm,ab}(\hat \Omega ),$ and 
$B_{jm,ab}(\hat \Omega )$
are eigenfunctions of ${\cal O}.$ Here the indices $a,b$ label
an orthonormal basis on the sphere. $T,$ and $E,$ and $B$ are represented 
as symmetric second-rank tensors on the sphere.  
Our task is to evaluate these eigenvalues, 
and following the notation for the temperature case, the eigenvalue may be expressed as 
$\frac{1}{2}{\cal N}_{white}^{-1}w^P_j$ where as before $w^T_j,~w^P_j\to 1$
as $j\to \infty .$ The temperature eigenvalues have already been calculated above, so we 
consider only the polarization eigenvalues, which are equal for $E$ and $B.$ As in the 
temperature case, for fixed $j,$ we may choose any point as a reference point and any 
convenient value of $m$
for calculating this eigenvalue. It turns out that the expression is the simplest
for $m=2$ with the reference point taken to be the north pole. 
[The only values of $m$ for which $\mathbf{E}_{jm}$ and likewise $\mathbf{B}_{jm}$
do not have zeros at the poles of the celestial sphere are $m=\pm 2.$]
Our starting point is the following equation
\begin{eqnarray}
\sum _{(i)}
\int _{-\pi }^{+\pi }\frac{d\bar \phi }{2\pi }~
\int _{-\infty }^{+\infty }dT~g(T)~
\mathbf{E}_{j2}
\Bigl(
\theta _{(i)}(T, \bar \phi ),
\phi _{(i)}(T, \bar \phi )
\Bigr)
\cdot \mathbf{e}(T;+, \bar \phi )
={\mathcal{N}_{white}}^{-1}~w_j^P~
\mathbf{E}_{j2}(N.P.)\cdot
\mathbf{e}_{+,N.P.}
\label{PolEVeqn}
\end{eqnarray}
which is the analogue for the polarization of eqns.~(\ref{evEqnTemp})
and (\ref{evEqnTempB}) for the temperature case.
We highlight some of the differences. Since the eigenvalue equation for
the polarized case has two components, we single out one component
of the polarization: namely, the component corresponding to the real
part of the spherical harmonic $\mathbf{E}_{j2}$ at the north pole,
which corresponds (up to a positive real multiplative factor) to
$
(
\hat {\mathbf{e}}_x\otimes \hat {\mathbf{e}}_x
-
\hat {\mathbf{e}}_y\otimes \hat {\mathbf{e}}_y)
/\sqrt{2}
$
in the ambient three-dimensional space into which the two-sphere is
embedded, or $
\cos (2\phi )
(\hat {\boldsymbol{\theta }}\otimes \hat {\boldsymbol{\theta }}-
\hat {\boldsymbol{\phi    }}\otimes \hat {\boldsymbol{\phi   }})/\sqrt{2}
+
\sin (2\phi )
(\hat {\boldsymbol{\theta }}\otimes \hat {\boldsymbol{\phi    }}+
\hat {\boldsymbol{\phi    }}\otimes \hat {\boldsymbol{\theta  }})/\sqrt{2}
$ in the (singular) usual spherical zweibein basis. Unlike in the temperature
case (because of the lack of invariance of the eigenfunction under rotations about
the $\hat {\mathbf{z}}$-axis), we have had to add another continuous label $\bar \phi $ 
to the paths passing through the reference point at the north pole. Here a path 
in the family $\bar \phi $ is at $T=0$ is directed toward 
$ \cos \bar \phi ~\hat {\boldsymbol{\theta }} + \sin  \bar \phi ~\hat {\boldsymbol{\phi }}.$
The unit vector indicating a polarization orientation
$\mathbf{e}_{(i)}(T;+, \bar \phi )$ corresponds to the polarization measurement at time $T$ 
along the path labeled by $(i)$ and $\bar \phi .$ At $T=0$ its orientation coincides with  
$\mathbf{E}_{j2}(N.P.)$ and at other $T$ it follows the path of the orientation of the polarization
measurement through the scan. 

We may rewrite the integrand of the integral over $\bar \phi $ on the left-hand side of 
eqn.~(\ref{PolEVeqn}) so that there is only one path from each class $(i)$ that at at $T=0$
points in the $\hat {\mathbf{x}}$ (or $\phi =0$) direction 
by exploiting the fact that upon a rotation by $\bar \phi $ along the 
$\bar {\mathbf{z}}$ direction,
the spherical harmonic changes by a factor $\exp [2i\bar \phi ].$ 
Eqn.~(\ref{PolEVeqn}) may thus be rewritten to become
\begin{eqnarray}
&&
\sum _{(i)} f_{(i)}
\int _{-\pi }^{+\pi }\frac{d\bar \phi}{2\pi } ~
\int _{-\infty }^{+\infty }dT~g(T)~
\textrm{Re}\left[ \exp [-2i\bar \phi ] ~
\mathbf{E}_{j2} 
\Bigl(
\theta _{(i)}(T, \bar \phi =0),
\phi _{(i)}(T, \bar \phi =0)
\Bigr) 
\right] 
\cdot \mathbf{e}(T;+, \bar \phi =0) \cr 
&&\qquad \qquad =
w_j^P \textrm{Re}\left[
\phantom{\bigg|}
\mathbf{E}_{j2}(N.P.)\cdot \mathbf{e}_{+,N.P.}
\right] . \phantom{\Bigg|} 
\end{eqnarray}
The explicit form of the mode functions is given in the Appendix. 
Here rather than rotating the path, we rotate the tensor spherical harmonic.
We adopt a notation singular to that used for spin-weighted
spherical harmonics (see \cite{spinWeighted} for more
details) where the $\hat {\mathbf{e}}_+$ component
is represented as the real part and
the $\hat {\mathbf{e}}_\times $ component as the
imaginary part of a complex function (or number) to
represent the polarization tensor on the celestial
sphere. The real part of $\mathbf{E}_{j2}$ rotated by an
angle of $-\bar \phi $ about the $\hat {\mathbf{z}}$
direction is thus represented as
\begin{eqnarray}
\cos [2(\phi +\bar \phi )]~Q_{j,2}(\cos \theta _i(T)) 
+i\sin [2(\phi+\bar \phi )]~U_{j,2}(\cos \theta _{(i)}(T)) .
\end{eqnarray}
We obtain
\begin{eqnarray}
w^P_j&=&1-\sum _{(i)}~f_{(i)}
\int _{-\infty }^{+\infty }dT~g(T)~
\int _{-\pi }^{+\pi }\frac{d\bar \phi}{2\pi } ~
\frac{1}{Q_{j,2}(1)}
\cr 
&&\qquad \times
\textrm{Re}
\Biggl[
\Bigl( \cos [2(\phi _{(i)}+\bar \phi )]~Q_{j,2}(\cos \theta _i(T))
+i\sin [2(\phi _{(i)}+\bar \phi )]~U_{j,2}(\cos \theta _{(i)}(T))\Bigr)\cr
&&
\qquad
\qquad
\qquad
\qquad
\times 
\exp \Bigl[ 
-2i\bigl\{ \bar \phi +\Delta \chi _{(i)}(T)\bigr\} 
\Bigr] ~
\Biggr]  
\cr &=&
1-\sum _{(i)}~f_{(i)}
\int _{-\infty }^{+\infty }dT~g(T)~
\frac{
Q_{j,2}(\cos \theta _i(T))+
U_{j,2}(\cos \theta _i(T))
}
{2Q_{j,2}(1)}
\exp \Bigl[ 
2i\bigl\{ \phi _{(i)}(T)-\Delta \chi _{(i)}(T)\bigr\} 
\Bigr] \cr
&=&
1-\sum _{(i)}~f_{(i)}
\int _{-\infty }^{+\infty }dT~g(T)~
\frac{
Q_{j,2}(\cos \theta _i(T))+
U_{j,2}(\cos \theta _i(T))
}
{2Q_{j,2}(1)}
\cos \Bigl[
2\bigl\{ \phi _{(i)}(T)-\Delta \chi _{(i)}(T)\bigr\}
\Bigr] .
\label{wPol}
\end{eqnarray}
Here $\phi _{(i)}(T)$ represents the azimuthal angle of the 
$(i)$ scan with $\bar \phi =0$ and $\Delta \chi _{(i)}(T)$ indicates the evolution
of the polarization measurement relative to the 
$(\hat {\boldsymbol{\theta}}, \hat {\boldsymbol{\phi }})$ 
zweibein and the initial orientation of the measurement at $T=0.$
Because of the hypothesis of nonchirality, the scans labelled by $(i)$
occur in pairs $\pm \phi (T)$ and $\pm \Delta \chi (T),$ 
allowing us to replace the exponential with the cosine in going 
to the last line of the above equation. 
Note that $Q_{j,2}(1) = U_{j,2}(1).$
In the above derivation, we have assumed that the initial
orientation of the polarizer for each $\bar \phi$ is aligned with
the polarization orientation of the real part of the $\mathbf{E}_{j2}$
tensor harmonic at the north pole reference point, the other cross polarization
giving zero. We are allowed to do this on account of a more general property
that we now state. Measuring one polarization orientation 
half the time and another polarization orientation 
rotated at $45^\circ $ the other half the time 
is equivalent to measuring polarization orientations
uniformly distributed on the circle. Here we refer to the initial orientation at $T=0$,
the subsequent orientations along the scan evolving with time according to $\Delta \chi _{(i)}(t).$
We may even relax the requirement on the distribution of the initial polarization 
orientation $\bar \chi $ further, requiring only that
$\left< \sin (4\bar \chi )\right> =\left< \cos (4\bar \chi )\right> $ and 
$\left< \sin (2\bar \chi ) \cos (2\bar \chi )\right> =0.$

\section{Numerical results: polarized case}

\begin{figure}
\vskip -0.15in 
\begin{center}
\includegraphics[width=0.4\textwidth]{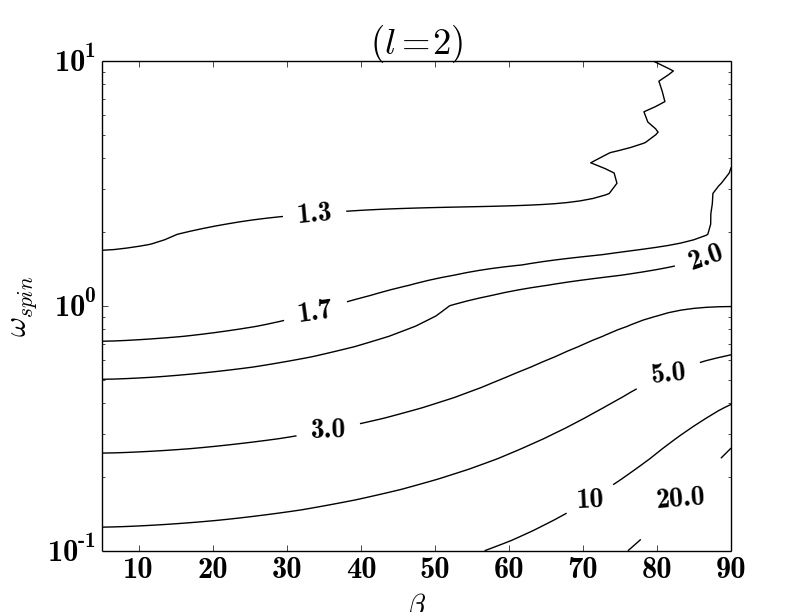}\\
\includegraphics[width=0.4\textwidth]{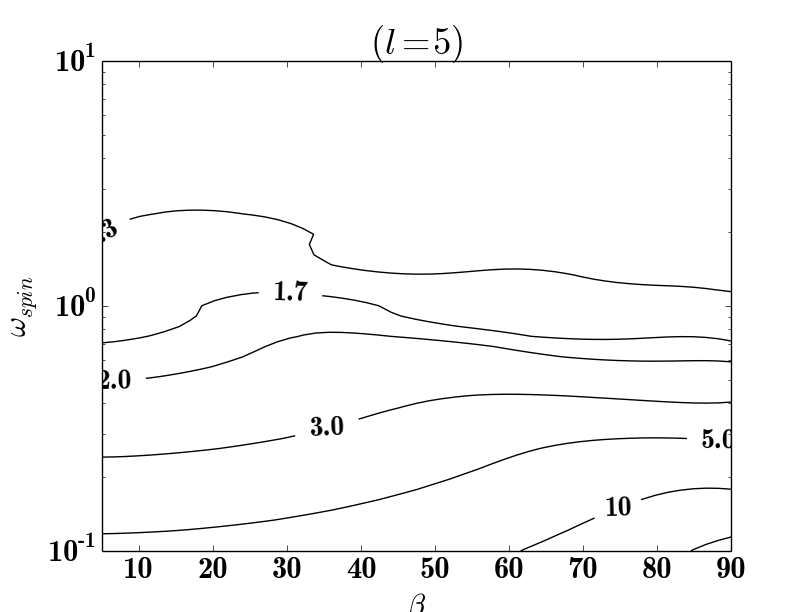}\\ 
\includegraphics[width=0.4\textwidth]{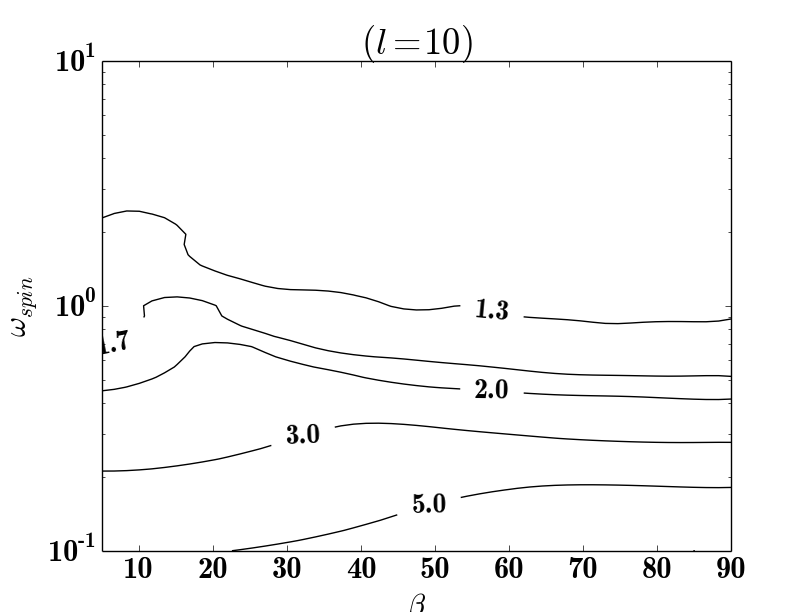}\\
\end{center}
\caption{\label{numFigPolarNoHWP} {\bf Polarization $1/f$ noise enhancement factor (without 
rotating HWP).}
In a manner analogous to Fig.~\ref{numFig}, we show the polarization boost factor $(w_\ell ^P)^{-1}$ for a few 
representative values of the multipole number $\ell .$
}
\end{figure}

We present some numerical results for the polarized case, first considering
the case with no rotating HWP. In this case two relevant parameters determine the 
$1/f$ noise boost factor: the circle opening angle $\beta $ and the 
ratio of the spin angular velocity to the $1/f$ knee frequency, expressed as the 
dimensionless ratio $\omega _{spin}=\Omega _{spin}/(2\pi \nu _{knee}).$
The scan trajectories are as considered in Section \ref{SectNumTemp}
for the temperature case, but here we additionally have to keep track of
the evolution of the azimuthal angle $\phi _{(i)}(t)$ and the orientation of the 
linear polarizer $\chi _{(i)}(t).$ Therefore we first compute explicit expressions for these
two quantities for the scan circles of the worked example.

\begin{figure}
\includegraphics[width=0.6\textwidth]
{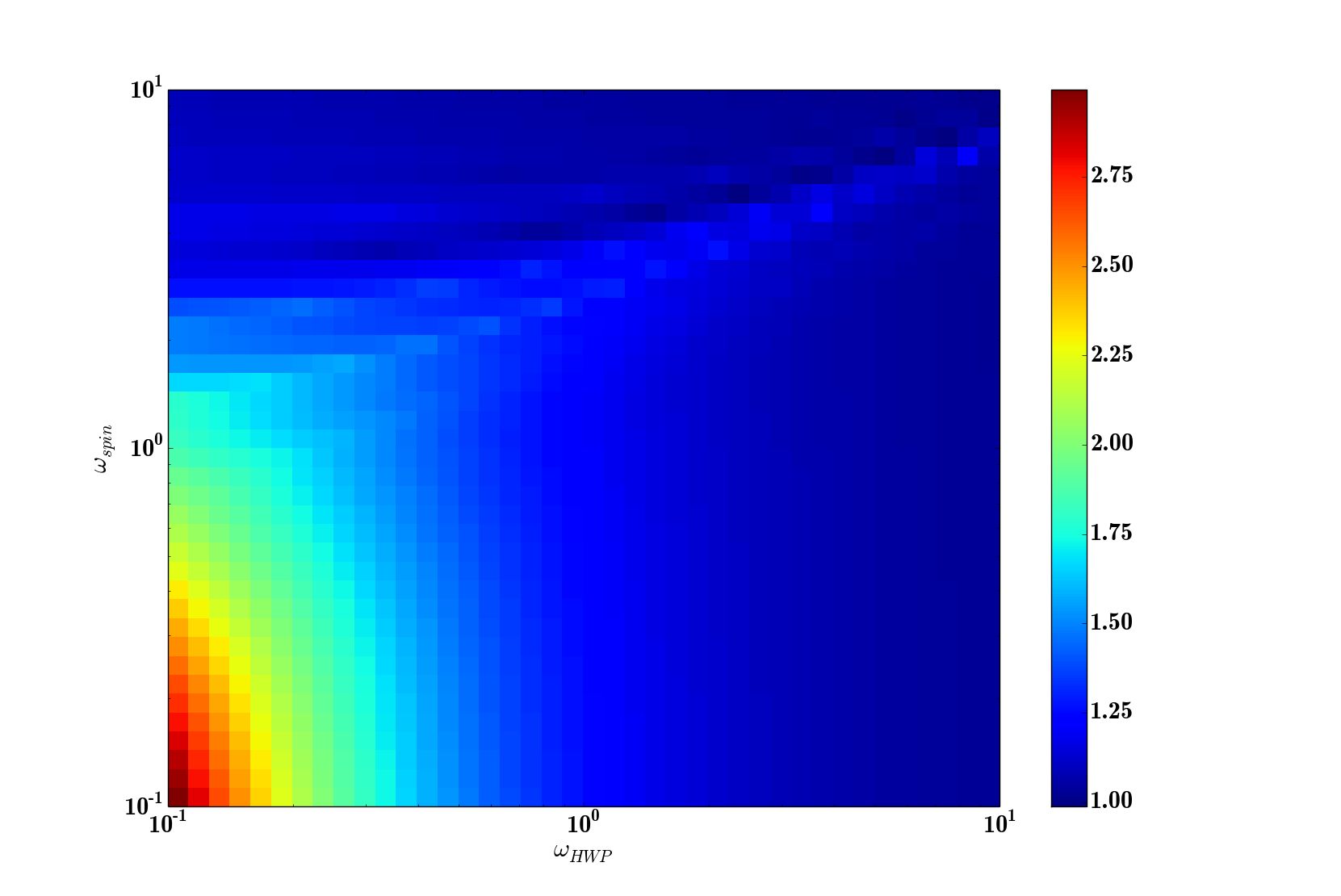}
\caption{\label{numFigPolarRotatingHWP} {\bf Polarization $1/f$ noise enhancement factor 
(with rotating HWP).} 
For $\ell =2$ and $\beta =60^\circ $ we show the boost factor as a function of $\omega _{spin}$
and $\omega _{HWP}.$ Here the relative signs of these two angular velocities does matter,
and when both signs are positive, there is more modulation and hence a smaller boost factor.
}
\end{figure}

For the rigidly spinning satellite taken as the worked example,
it is convenient to specify the direction of the polarizer relative to the scan direction,
as in this case unless there is a rotating half-wave plate, the relative orientation
of the two vectors remains constant. The instantaneous scan direction is given by 
\begin{eqnarray}
\hat {\mathbf{v}}
=
\frac{\dot {\hat {\mathbf{n}}}(t)}{\vert \vert \dot {\hat {\mathbf{n}}}(t) \vert \vert }.
\label{ScanDirection}
\end{eqnarray}
Let $\hat {v}_\theta (t)$ and $\hat {v}_\phi (t)$
denote the components of 
$\hat {\mathbf{v}}$ with respect to the orthonormal basis
$\hat {\boldsymbol{\theta }},$
$\hat {\boldsymbol{\phi }}.$
Then we may write
\begin{eqnarray}
\exp \Bigl[ 2i\chi (t)\Bigr] =
\left(
\frac{
\hat {v}_\theta (t)+i\hat {v}_\phi (t)
}{
{\hat {v}_\theta }^2(t)+{\hat {v}_\phi }^2(t)
}
\right) 
\exp \Bigl[ 2i\delta (t)\Bigr] 
\end{eqnarray}
where $\delta (t)$ indicates the angle of polarizers relative to the scan direction 
$\hat {\mathbf{v}}.$ In the worked example, 
$\delta (t)=(\textrm{constant}).$
We now compute $\hat {\mathbf{v}}(t)$ as defined in eqn.~(\ref{ScanDirection}). 
Taking the time derivative of explicit trajectory for the worked example
\begin{equation}
\begin{pmatrix}
\hat n_x(t)\\
\hat n_y(t)\\
\hat n_z(t)
\end{pmatrix}
=
\begin{pmatrix*}[l]
\sin \beta ~\sin (\Omega _{spin}t)\\
\sin \beta ~\cos \beta \bigl( \cos (\Omega _{spin}t)-1\bigr) \\
\sin ^2 \beta \cos (\Omega _{spin}t)+\cos ^2\beta 
\end{pmatrix*}, 
\label{spinPatternBis}
\end{equation}
which was derived in eqn.~(\ref{spinPattern}) above, we obtain 
\begin{eqnarray}
\begin{pmatrix}
{\dot {\hat n}_x}(\tau )\\
{\dot {\hat n}_y}(\tau )\\
{\dot {\hat n}_z}(\tau )
\end{pmatrix}
=
\Omega _{spin} \sin \beta 
\begin{pmatrix*}[l]
\phantom{-\sin \beta }\cos (\Omega _{spin}t )\\
-\cos \beta \sin (\Omega _{spin}t ) \\
-\sin \beta \sin  (\Omega _{spin}t )
\end{pmatrix*},
\end{eqnarray}
and
\begin{eqnarray}
\hat {\mathbf{v}}(t)
=
\begin{pmatrix*}[l]
{\hat v}_x(t)\\
{\hat v}_y(t)\\
{\hat v}_z(t)
\end{pmatrix*}
=
\begin{pmatrix*}[l]
\phantom{-\sin \beta }\cos (\Omega _{spin}t )\\
-\cos \beta \sin (\Omega _{spin}t ) \\
-\sin \beta \sin  (\Omega _{spin}t )
\end{pmatrix*}.
\end{eqnarray}
Using the basis expressed in the three-dimensional ambient coordinates
(of the space into which the sphere is embedded)
\begin{eqnarray}
\hat {\boldsymbol{\theta }}(t)&=&
\begin{pmatrix*}[l]
{\hat \theta }_x(t)\\
{\hat \theta }_y(t)\\
{\hat \theta }_z(t)\\
\end{pmatrix*}
=
\begin{pmatrix*}[l]
+\cos \theta (t)~\cos \phi (t)\\
+\cos \theta (t)~\sin \phi (t)\\
-\sin \theta (t)
\end{pmatrix*}
=
\begin{pmatrix*}[l]
\hat n_x(t)~ \hat n_z(t)\Big/ \sqrt{1-{\hat n_z(t)}^2}\\
\hat n_y(t)~ \hat n_z(t)\Big/ \sqrt{1-{\hat n_z(t)}^2}\\
-\sqrt{1-{\hat n_z(t)}^2}
\end{pmatrix*},\cr
\hat {\boldsymbol{\phi }}(t)&=&
\begin{pmatrix*}[l]
{\hat \phi }_x(t)\\
{\hat \phi }_y(t)\\
{\hat \phi }_z(t)\\
\end{pmatrix*}
=
\begin{pmatrix*}[r]
-\sin \phi (t)\\
+\cos \phi (t)\\
0
\end{pmatrix*}
=
\begin{pmatrix*}[l]
-\hat n_y(t)/\sqrt{1-{\hat n_z(t)}^2}\\
+\hat n_x(t)/\sqrt{1-{\hat n_z(t)}^2}\\
0
\end{pmatrix*},
\end{eqnarray}
where $\hat n_x(t),$ $\hat n_y(t),$ and $\hat n_z(t)$
are as defined in eqn.~(\ref{spinPatternBis}),
we obtain 
\begin{eqnarray}
\hat {v}_\theta (t)&=& 
\hat {\theta }_x(t)~{\hat v}_x(t)+
\hat {\theta }_y(t)~{\hat v}_y(t)+
\hat {\theta }_z(t)~{\hat v}_z(t)
,\cr 
\hat {v}_\phi (t)&=& 
\hat {\phi }_x(t)~{\hat v}_x(t)+
\hat {\phi }_y(t)~{\hat v}_y(t)+
\hat {\phi }_z(t)~{\hat v}_z(t).
\end{eqnarray}
We also have 
\begin{eqnarray}
\exp \Bigl[ 2i\phi (t)\Bigr] =
\frac{{\hat n}_x(t)+i{\hat n}_y(t)}{{\hat n}_x(t)-i{\hat n}_y(t)}.
\end{eqnarray}
Similarly we obtain
\begin{equation}
\exp [ 2i\Delta \chi ]=
\frac{\hat v_\theta +i\hat v_\phi }
{\hat v_\theta -i\hat v_\phi },
\end{equation}
so that we now have all the ingredients necessary
to evaluate eqn.~(\ref{wPol}) for the worked example. For the case of a rotating HWP,
the above expression is replaced with 
\begin{equation}
\exp [ 2i\Delta \chi ]=
\frac{\hat v_\theta +i\hat v_\phi }
{\hat v_\theta -i\hat v_\phi }~\exp[-4i\Omega _{HWP}].
\end{equation}
Here the coordinates have been chosen such that the scan moves along the $\phi =0$ direction for $t=0+.$ 
The trajectory of the scan is given by 
$ \bigl( \theta _{(i)}(t), \phi _{(i)}(t) \bigr),$ and  
$\chi _i$ is the orientation 
of the polarizers using the 
$\hat {\boldsymbol{\theta }},$ 
$\hat {\boldsymbol{\phi }}$ 
basis. 
$\Omega _{HWP}$ is the HWP angular rotation velocity.

The expression in eqn.~(\ref{wPol}) for the polarized noise differs from the expression in 
eqn.~(\ref{wTemp}) for the unpolarized noise in one important respect apart from
the precise form of the mode functions $Q_{j2}(\cos \theta )$ and 
$U_{j2} (\cos \theta )$ as compared to the Legendre polynomial $P_j(\cos \theta ).$ 
The most significant difference is 
that for small multipole number $j,$ while in the temperature integral in eqn.~(\ref{wTemp}) 
the only way to make this integral small compared to one is to move the beam far away 
from the reference point before a time of order ${\nu _{knee}}^{-1}$ has elapsed, in the 
integral for the polarized case it is possible to obtain cancellations through the oscillations 
in the cosine factor, a multiplicative factor that does not occur for the temperature case.
Qualitatively, this feature of the polarized integral means that $1/f$ excess can be eliminated 
without scanning over large circles provided that the HWP rotates faster than $\nu _{knee},$ or 
alternatively if the orientation of the polarization $\chi _{(i)}(t)$ changes quickly compared to 
$\nu _{knee}.$ For a satellite spinning at a rate $\Omega _{spin}$ with a detector at boresight 
angle $\beta $ where $\beta $ is small, the polarization angle rotates at approximately the 
same rate assuming that the assembly rotates as a rigid body. For this case we 
assume no additional mechanical or other means of rotating the polarization direction 
of the detector. We can also consider the case where there is additionally a rotating
HWP in front of the detector or the telescope. 
Within the framework of the 
calculations above, rotating the HWP is equivalent to rotating the detector about its 
optical axis because here we do not consider nonideal beams that lack azimuthal symmetry
(in the absence of a precise correction 
allowing the local second-derivative in the $T$ anisotropy to masquerade as a polarized signal).

For the polarized case, the classification of ways of crossing the reference pixel,
denoted here using the sum $\sum _{(i)}f_{(i)},$ should in principle 
also include the direction of the 
polarizer with respect to the scan direction, thus adding another continuous angular variable
$\bar \chi $ to the classification.  
When there is a rotating half-wave plate, its effect is simply to rotate the direction of the 
measured linear
polarization on the sky, so its orientation at reference pixel crossing can be absorbed into the
variable $\bar \chi .$
However, because the polarizer orientation enters into the final expression in the form of the sum 
of a term multiplied by $\cos (2\bar \chi )$ and another term multiplied by $\sin (2\bar \chi ),$ it is
possible to avoid distinguishing the polarizer orientations using a continuous classification.
Instead one can consider only four orientations at $\bar \chi =0, \pi /4, {\pi }/{2},$ and 
${3\pi }/{4},$ it being understand that $\bar \chi $ and $(\bar \chi +\pi )$ correspond to the same 
orientation.  

This last property also enlarges the class of scan patterns that may be analyzed as being isotropic. 
Any survey that measures a polarization direction and also the polarization 
direction rotated by $\pm 90 ^\circ $ over the same
fraction of scans and for which the combinations 
$+$ and $\times $ are covered equally can be analyzed 
exactly the formulas as for the isotropic case.
More generally, these conditions may be expressed 
as 
$\left< \cos 4\bar \chi \right>= \left< \sin 4\bar \chi \right> $
and 
$\left< \cos 2\bar \chi \sin 2\bar \chi \right> =0.$
In other words, when these two conditions are satisfied, 
all the formulae derived in this paper apply, even though strictly
speaking, isotropy might not respected by the distribution of polarizer orientations. 

The results without a rotating half-wave plate are shown in Fig.~\ref{numFigPolarNoHWP}.
Compared to the temperature case, 
the most significant qualitative difference is that when $\beta $ 
is small but $\omega _{spin}$ is large, considerable suppression of $1/f$ noise boost 
is obtained. This situation contrasts with the temperature case, where small $\beta $ 
is always accompanied by a large $1/f$ boost factor at low multipole number no matter how
large $\omega _{spin}$ is.

Although we shall not provide any examples here, others schemes for suppressing $1/f$ noise without 
resorting to a rotating HWP can be envisaged. For example, one might rotate the polarization 
direction as the detector moves in a way different from the scheme above where we have assumed 
the satellite spinning rigidly together with a rigidly attached detector assembly. Rotating the 
detector or the telescope relative to the spinning satellite might introduce additional 
technical complications but could provide an alternative to a rotating HWP.
We now consider a rotating HWP. This case involves an extra dimensionless parameter: $\omega _{HWP}
=\Omega _{HWP}/(2\pi \nu _{knee}).$ These results are shown in Fig.~\ref{numFigPolarRotatingHWP}.
We observe that a large $1/f$ boost at low multipole can be avoided for the polarization even
when $\omega _{spin}$ is small.

\begin{figure}[h]
\includegraphics[width=3in]
{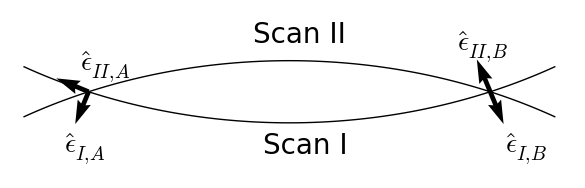}
\caption{\label{flatSkyDiff} 
{\bf Measuring flat sky polarization differences.}  
The linear polarizations are detected at a constant angle with 
respect to the two arcs labelled Scan I and Scan II, so that
at the points of intersection, labelled $A$ and $B,$ measurements
are made along 
$\hat {\boldsymbol{\epsilon}}_{I,A},$
$\hat {\boldsymbol{\epsilon}}_{I,B},$
$\hat {\boldsymbol{\epsilon}}_{II,A},$ and 
$\hat {\boldsymbol{\epsilon}}_{II,B},$
respectively.
$\hat {\boldsymbol{\epsilon}}_{I,B}$ and $\hat {\boldsymbol{\epsilon}}_{II,B}$ are anti-parallel
while 
$\hat {\boldsymbol{\epsilon}}_{I,A}$ and $\hat {\boldsymbol{\epsilon}}_{II,A}$ are orthogonal.
As explained in the main text, this is how the polarization orientation is naturally carried 
across the curved sky for a spinning satellite without a rotating half-wave plate.
In this way, the difference in intensity at point $A$ between the polarizations 
$\hat {\boldsymbol{\epsilon}}_{I,A}$ and $\hat {\boldsymbol{\epsilon}}_{II,A}$
can be expressed using solely difference along the same scan, avoiding the need to measure 
the offsets of the two scans.
}
\end{figure}

\section{Flat sky approximation}

The above main results, in particular eqns.~(\ref{wTemp}) and (\ref{wPol}), readily generalize 
to a flat sky. The Legendre polynomials $P_\ell (\cos \theta )$ are replaced with 
the spherical Bessel functions $j_0(kr)$ 
where the discrete variable $\ell $ becomes replaced with the continuous variable $k,$
and the $E$ mode basis functions are generated by the second derivative operator acting on $j_2(kr) \exp 
[2i\phi ]$ with its trace removed.
Some of the qualitative results discussed earlier in this paper can be understood very simply 
in the flat sky approximation, for example how the contribution of $1/f$ noise to $U$ and $Q$ 
can be removed either by using small circles or by a rotating HWP (provided that the spin rate 
or the HWP rotation rate, respectively, is greater than $\nu _{knee}).$ Figure 
\ref{flatSkyDiff} shows scan circles intersecting at $45^\circ .$ Now we can compare along scan 
I, the polarizations at $A$ and $B$ as indicated, and then $B'$ and $A'$ along scan II. In this 
way one linear combination of $Q$ and $U$ can be expressed in terms of differences of 
measurements along the same scan.
Let us examine a scan pattern that cannot measure a sky polarization that is 
constant or with a very small wave number. 
(Of course, a constant polarization is not possible on a curved sky 
but is possible in the flat sky approximation.) 
Suppose we consider scans along straight lines with no rotating HWP. 
(This scan pattern would be somewhat analogous to scanning along great circles on the curved 
sky.\footnote{In this case, however, because of the curvature, for example
by combining three great circles 
that form a triangle whose interior angles are all right angles (forming a sort of an eighth of a 
orange, which has been halved three times), 
the polarization in a certain direction and the polarization in a direction at 
a right angle to the first direction can be compared by considering only differences of 
measurements along the same scan.}) In this case crossing scans do not connect orthogonal 
polarizations. Consequently there is no way to constrain the difference between the unknown 
zero point offsets of orthogonal polarizations.

\section{Relevance to actual surveys}

The calculations of the previous Sections pertain to an idealized `isotropic' survey. Yet 
this assumption is not respected for actual surveys, primarily because of the need to avoid
pointing close to the Sun. We first give an overview of the scanning patterns of the three
space missions that have already flown (i.e., COBE, WMAP, and Planck) and then describe a
family of scanning patterns being considered for future surveys. More precise numerical
studies will be the subject of a future publication.

\subsection{The COBE, WMAP, and Planck scanning patterns}

We briefly review the scanning
patterns of the three CMB satellite experiments that have
already flown with an emphasis on elucidating the relevance,
or perhaps lack thereof, of the methods developed in this 
paper to the map making problem for these missions. We
do not consider ground and balloon based experiments 
because such experiments observe 
fields on the sky with boundaries. Their scanning patterns moreover are
constrained to follow 
circles at constant angle to the zenith 
because of the large gradient in sky temperature
arsing from atmospheric emission. 

To understand these three satellite experiments, a
distinction can be drawn between the so-called `differential power'
instruments, such as COBE and WMAP, 
and the so-called `total power' 
instruments, such as the Planck HFI instrument and all
the fourth-generation CMB satellites that have been
proposed (i.e., SAMPAN, EPIC, B-Pol, COrE, PRISM, COrE+, and
LiteBird) with the sole exception of PIXIE. 
Differential measurement instruments such 
as COBE and WMAP have paired horn assemblies or telescopes,
respectively, pointing at widely separated locations 
on the sky, and differences are taken by switching
a detector between a pair of horns recording
only the difference. The $1/f$ noise is thus 
almost completely eliminated, but at the expense 
of throwing away half the data. For coherent detectors,
for which the $1/f$ noise time scale is very short, 
such switching is unavoidable.\footnote{The Planck LFI 
instrument could perhaps formally be considered a 
total power instrument because there was no switching
between pairs of pixels on the sky, but the LFI
overcame the large $1/f$ noise inherent in contemporary 
coherent (HEMT) detectors by switching between 
the sky and a highly stable, internal cold reference load.}
Modern bolometers, however, have been able to achieve
the stability needed to render total power measurements
practical. The Planck HFI was a total power
instrument.
Total power measurements allow all data taken
to be exploited rather than just exploiting the differences
comprising only half the data.
They also simplify the design by avoiding the need for
two telescopes or pairs of horns on the sky,
as is required for making differential measurements.

We first briefly describe the Planck scan pattern, 
presented in detail in \cite{tauber}, which strongly
violates the hypothesis of isotropy, whose consequences 
have been explored in this 
paper. With its focal plane covered by 70 detectors, the Planck 
instrument sparsely samples an $8^\circ $ field of view.
Through the spin of the satellite, circles on the sky $\approx 85^\circ $ in 
diameter are scanned $\approx 39-65$ times before moving
on to the next circle. Under the just described `nominal'
scanning pattern, the 
angle to the Sun is kept constant with 
the spin axis orthogonal to the solar direction. The 
spin axis is stepped 2' every hour when the scan
circle is changed. The nominal scan pattern just described
has almost no cross linking, and in most places only at a small
angle; therefore, two variations on this scan pattern were considered
to remedy this defect: a `cycloidal' precession and a `sinusoidal'
precession of the spin axis relative to the antisolar direction.  
For a cycloidal precession the angle of the spin axis with respect to the antisolar direction is kept constant. 
This means that when averaging over the spin rotation (which is relatively fast) the thermal state does
not vary. Under the alternative sinusoidal precession, the spin axis rises and falls sinusoidal 
relative to the ecliptic equator, but the motion parallel to the equator remains at a constant angular velocity, resulting
in more evenly spaced scanning circles. This advantage comes at the expense of varying angle with respect
to the antisolar direction, which in principle could introduce thermal artefacts. 
In the end, a cycloidal
precession with the ecliptic latitude oscillating between $-7.5^\circ $
and $+7.5^\circ $ with a six-month period 
was used for the Planck scientific data taking.

With respect to the analysis methods developed in this paper,
that Planck scanning pattern is about as anisotropic as possible.
The sky coverage is not uniform, being densest around the ecliptic
poles and sparsest around the ecliptic equator. Moreover there is
little cross linking except near the poles. These aspects of the Planck
scanning pattern have been criticized by some. Nevertheless the Planck
scanning pattern is not without its advantages, the most noteworthy of
which is the large number of redundant measurements around a scanning
circle, allowing for a precise characterization of the noise properties
of the detectors and the evolution of these properties in time.

The WMAP scanning pattern is described in detail in refs.~\cite{wmapBasicResults}
and \cite{pageOpticalDesign}. 
The WMAP dual telescopes point at $\pm 70^\circ $
relative to the satellite spin axis (the exact value
depending slightly on the precise detector pair), and this spin
axis in turn precesses at an angle of $22.5^\circ $
relative to the antisolar direction, so that over
an intermediate timescale of a few hours to a few days
an annular region of inner radius $\approx 47.5^\circ $
and outer radius $\approx 92.5^\circ $ is 
mapped. The spin frequency
was 0.464 revolutions per minute while the precession
rate was one revolution per hour.
Owing to the annual motion of the antisolar direction on
the celestial sphere, this annulus moves around the
ecliptic equator, so that the whole sky is mapped
approximately every six months.
Because of the rapid switching between detectors, the time stream of differences
in temperature between paired sky pixels was almost without $1/f$ noise, and it was
possible to make sky maps using a simple weighted least squares method, which
is computationally much less demanding than the more sophisticated map making
techniques taking into account $1/f$ of the type described above.
Moreover the large degree of cross linking of the WMAP scan pattern was helpful for 
measuring the polarized signal. For the WMAP analysis, one of the greatest challenges was 
characterizing the gain variations of the detectors, a complication not explored in this paper.

For the COBE satellite, which was in low-Earth orbit unlike WMAP and Planck, which where placed
at L2, the spin axis pointed almost exactly in the anti-Earth direction. The COBE satellite 
orbited the Earth with a period of approximately 90 minutes, its orbit carefully placed in the 
`twilight zone'---that is, in the plane at right angles to the solar direction. 
The optical axis of each horn subtended an angle of approximately $30^\circ $ with 
respect to the spin axis of the satellite, whose spin rate was approximately 0.8 rpm. 

Both the WMAP and COBE scan patterns had substantially greater cross linking than
the Planck scan pattern and covered the sky in a more uniform manner, closer 
to isotropic. 

\subsection{Spin-precession family of scanning patterns}

\begin{figure}
\begin{center}
\includegraphics[width=5in]{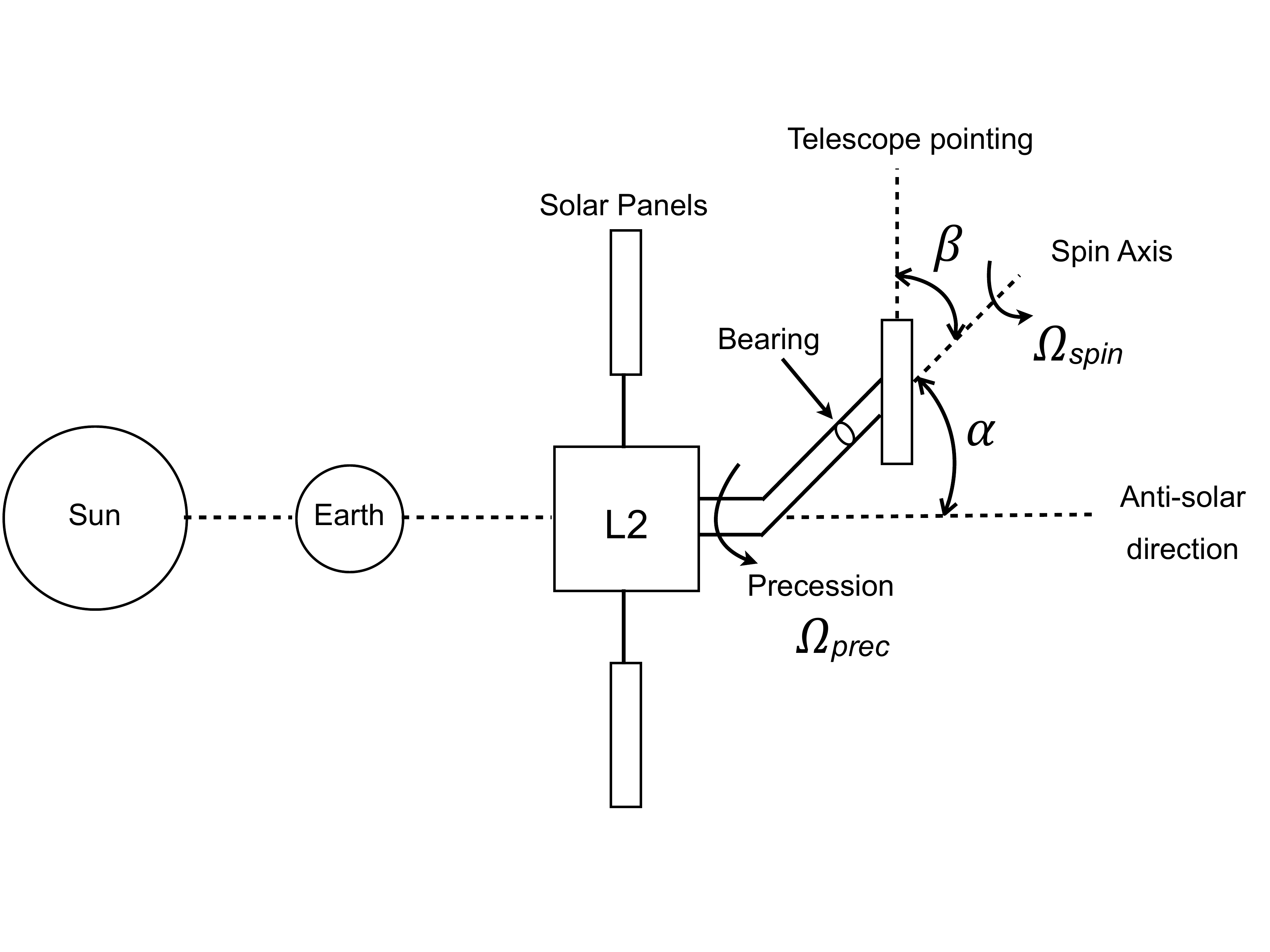}
\end{center}
\caption{
{\bf Satellite setup for a spin-precession scan pattern.} 
The relative positions of the Sun, Earth, and the satellite (located at L2)
are shown. The precession axis of the satellite is aligned with 
the antisolar direction so that the antenna for downloading data
always points toward the Earth and the solar panels always lie in the 
plane perpendicular to the antisolar direction. The spin axis
subtends an angle $\alpha $ with the antisolar direction and by
means of a bearing the telescope is allowed to rotate about this 
spin axis. The optical axis of the telescope subtends an angle
$\beta $ with the spin axis.
}
\label{SpinPrecTwo}
\end{figure}

This Subsection describes what we shall call the `spin-precession' family of scan patterns. Consider 
the setup sketched in Fig.~\ref{SpinPrecTwo}. A scan pattern belonging to this family
is envisaged for the present scanning configuration of the LiteBird satellite, which is 
to be placed at $L2,$ the second Lagrange point at which the satellite 
orbits behind the Earth and 
remains collinear with the Sun and the Earth at all times, 
as indicated in the Figure. The main axis of the satellite
always points precisely in the antisolar direction, so that the antenna 
used 
for downloading data is always directed toward the Earth 
and the solar panels always lie precisely in the plane normal to the solar 
direction. The telescope assembly is mounted obliquely on an axis at 
an angle $\alpha $ to the antisolar direction and a bearing allows the 
telescope assembly to spin about this axis, which we shall call the spin axis.
The optical axis of the telescope in turn subtends an angle $\beta $ with
the spin axis. The pointing of the telescope is subject to a hierarchy of
motions, the fastest of which is the rotation at angular velocity $\Omega _{spin}$
about the spin axis. The satellite assembly also precesses by rotating about
the antisolar direction at a rate $\Omega _{prec},$ and then finally there 
is the annual motion of the antisolar direction at the angular velocity $\Omega _{ann}.$

More complicated scan patterns could be envisaged under which pixels would be 
revisited more frequently and at crossing angles varying more between successive 
passes, a feature that could reduce artefacts arising from uncorrelated gain variation. 
Implementing such a scan pattern, however, would be technically more challenging,
requiring an additional level of modulation, or alternatively a completely free 
scan pattern controlled using inertial flywheels.

\begin{figure}
\begin{center}
\includegraphics[width=5in]{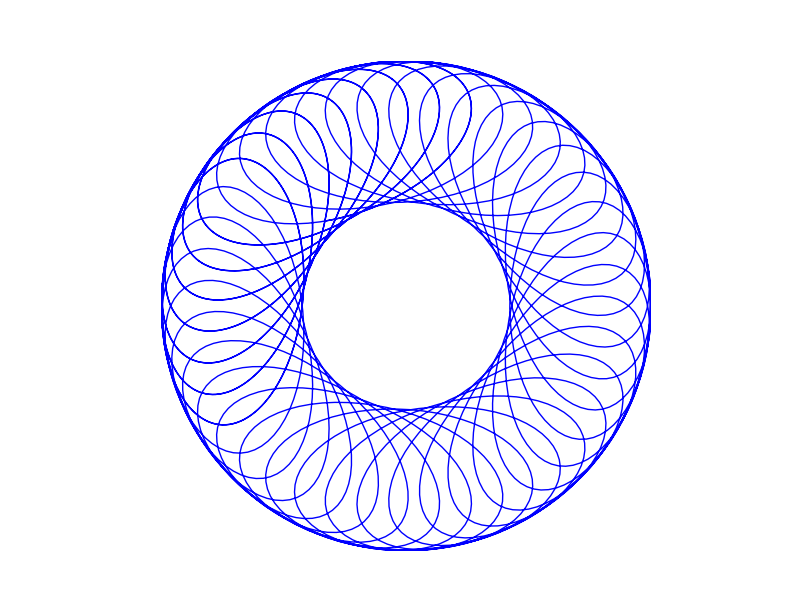}
\end{center}
\caption{
{\bf Spin-precession scan pattern.} We show
the scan pattern for $\alpha =60^\circ $ and $\beta =35^\circ $
with $\Omega _{prec}/\Omega _{spin}=1/30$ assuming 
$\Omega _{spin},\Omega _{prec}\ll \Omega _{ann}.$
The center of the figure corresponds to the antisolar direction, which
moves around the ecliptic plane over the course of the year.
}
\label{PrecPatternOne}
\end{figure}

The natural coordinate system for analyzing scanning patterns is
ecliptic coordinates because of the need to avoid pointing too close
to the Sun. We shall ignore here complications introduced by the
1.67\% eccentricity of the solar orbit by pretending that the solar
orbit is exactly circular. We define our coordinates so that $\hat {\mathbf{z}}$ 
points toward the north ecliptic pole, $\hat {\mathbf{x}}$ points in the 
antisolar direction at the vernal equinox, and 
$\hat {\mathbf{y}}= (\hat {\mathbf{z}} \times \hat {\mathbf{x}}).$
More explicitly, without the annual motion of the Earth around the Sun 
taken into account,
the orientation of the optical axis of the telescope is given by
\begin{eqnarray}
\hat {\mathbf{n}}_{opt}(t)
&=&
\bigl[ ~{\cal R}_{\hat {\mathbf{x}}}(\Omega _{prec}t)~
{\cal R}_{\hat {\mathbf{z}}}(\alpha )~
{\cal R}_{\hat {\mathbf{x}}}(-\Omega _{prec}t)~\bigr]
\bigl[ ~{\cal R}_{\hat {\mathbf{x}}}(\Omega _{spin}t)~
{\cal R}_{\hat {\mathbf{z}}}(\beta )~\bigr] ~
\hat {\mathbf{n}}_{x},
\end{eqnarray}
corresponding to a pattern as sketched in Fig.~\ref{PrecPatternOne}.
When the annual motion of the Earth around the sun is included, this expression
is modified to become
\begin{eqnarray}
\hat {\mathbf{n}}_{opt}(t)
&=&{\cal R}_{\hat {\mathbf{z}}}(\Omega _{ann}t)
\bigl[ ~{\cal R}_{\hat {\mathbf{x}}}(\Omega _{prec}t)~
{\cal R}_{\hat {\mathbf{z}}}(\alpha )~
{\cal R}_{\hat {\mathbf{x}}}(-\Omega _{prec}t)~\bigr] 
\bigl[ ~{\cal R}_{\hat {\mathbf{x}}}(\Omega _{spin}t)~
{\cal R}_{\hat {\mathbf{z}}}(\beta )~\bigr] ~
\hat {\mathbf{n}}_{x}.
\end{eqnarray}

We shall ignore the overall magnitude of the scan circle density 
and all issues concerning the small-scale structure of the scan pattern,
instead treating the density of circles as if it were a
smooth continuum density. We thus ignore
important issues 
beyond the scope of this article, whose primary aim
is to understand the impact of the $1/f$ noise,
which is most relevant on the largest angular scales.
To the extent that $\Omega _{prec}\ll \Omega _{spin},$ it is not inaccurate
to approximate the scan pattern using closed circles. This amounts to
replacing the continuous precession with a series of small discrete steps. 
In this approximation, without the motion of the Earth around the sun, the scan pattern
may be approximated as a series of circles of opening radius $\beta $ situated on the 
precession circle of radius $\alpha .$ The annual motion of the Earth around the sun
ensures that the distribution of circles on the celestial sphere is azimuthally symmetric.
Therefore to see how well isotropy is satisfied, it suffices to consider how 
the density of circles varies with $\theta .$
For the antisolar direction oriented in the $\hat {\mathbf{x}}$ direction,
the precession causes the instantaneous spin direction $\hat {\mathbf{n}}_{spin}(t)$ 
to evolve as 
\begin{eqnarray}
\hat {n}_{spin, x}=\cos \beta  ,\quad 
\hat {n}_{spin, y}=\sin \beta  ~\cos (\Omega _{prec}t),\quad 
\hat {n}_{spin, z}=\sin \beta  ~\sin (\Omega _{prec}t).
\end{eqnarray}
In the approximation where the precession is taken to be stepped rather than continuous,
this is the trajectory of the centers of the closed circles, which we may approximate to be
distributed continuously rather discretely. This pattern of course is rotated 
about the $\hat {\mathbf{z}}$ direction by 
the annual motion, which ensures azimuthal symmetry, so to check for isotropy
of the distribution of circles,
we need to consider only the distribution after projection along the $\hat {\mathbf{z}}$ direction.
Isotropy would correspond to a density uniform in $z$ extending from $z=-1$ to
$z=+1,$ but instead we obtain a distribution undersampling the equatorial regions
of the celestial sphere and oversampling the polar regions, with a mild singularity
either at the poles themselves, when $\beta =\pi /2,$ or at the edges
of the bald spots near the ecliptic poles, for $\beta <\pi /2.$ 
In the former case, $dN/dz=(N_c/2)(2/\pi )(1-z^2)^{1/2}$ rather than 
$dN/dz=(N_c/2)$ for the isotropic case. This difficulty could be mitigated 
by a nonuniform precession instead of the linear progression of the 
precession angle considered above in order to spend more time in the equatorial
region, and this softens the singularity at high latitude where the direction of
the precession in the direction normal to the ecliptic plane reverses sign. 
Nevertheless, one wants to keep $\beta $ large, which is necessary for 
minimizing the $1/f$ boost factor for the temperature, or $I$ Stokes parameter
noise, and at the same time keep $(\alpha +\beta ),$ the largest angle that the optical
axis subtends with the antisolar direction, not too large. This maximum angle
determines the requirements 
for shielding the telescope from the sun. This shielding requirement
is incompatible with a large value
of $(\alpha +\beta ),$ and when $(\alpha +\beta )< \pi /2,$ the distribution of scanning circles
on the celestial sphere has bald spots around the celestial sphere.
The bald spots in the distribution of circle centers
does not necessarily entail that the observations
do not extend to around the ecliptic 
poles, but rather that the pixels near the ecliptic poles are
not traversed isotropically.

The spin-precession family of scan patterns 
has the advantage that the mean angle
with respect to the antisolar direction after averaging over the spin rotation
(assumed fast) 
does not vary. This feature is helpful for minimizing systematic errors arising
from time varying thermal gradients induced within the instrument.
This family of scanning patterns 
also allows approximately half the sky to be surveyed
at any moment. On the other hand, even though with appropriately chosen $\alpha $
and $\beta ,$ pixels are revisited many times throughout a six month period,
the direction through which a given pixel is revisited varies slowly, changing
at a rate proportional to $\Omega _{ann}.$ This feature could be problematic
in the presence of time variations in the detector gain, especially for 
measuring polarization without a rotating half-wave plate.

\section{Concluding comments}

We close with the following observations:

(1) The main obstacle to implementing an isotropic scan pattern, at least one with circles of 
a large opening angle $\alpha ,$ is the presence of the sun, even for a satellite situated at 
the second Lagrange point $(L2),$ where avoiding the Earth and the Moon is less problematic. A 
circle with opening radius $\alpha $ not much smaller than $\pi /2$ whose center lies near the 
ecliptic poles cannot be surveyed at any time of the year without pointing very close to the 
sun. For a spinning satellite, the larger the maximum allowed angle of the boresight direction 
with the antisolar direction, the more drastic are the measures needed for adequate solar 
shielding, and also the larger the risk of thermal drifts at the spin frequency and its 
higher order harmonics. Of course, an isotropic scan pattern with smaller circles might not be 
too hard to implement, but it is not apparent that such an isotropic scan pattern would be 
superior to an anisotropic scan pattern with larger circles, at least for measuring the temperature
anisotropies.

(2) Using large circles is indispensable for reducing $1/f$ noise at low multipoles for 
measurements of the temperature anisotropy. However the same does not hold for measuring the 
polarization anisotropies, as we have demonstrated above. For small circles, the polarization 
$1/f$ noise excess at all multipoles can be suppressed by spinning the satellite sufficiently 
fast compared to the knee frequency $\nu _{knee},$ or alternatively by rotating a HWP considerably 
faster than $\nu _{knee}.$ For circles with $\beta \approx \pi /2$ (i.e., circles that are 
nearly great circles), the spin of the satellite helps significantly less to reduce the $1/f$ 
noise. In the case of exact great circles, if it were not for the curvature of the sky, one
would have two zero point offsets, one for each linear polarization, with no way of determining
their difference from the data taken in the survey.
For large great circles, however, a rotating HWP plate can be used to reduce the $1/f$ 
noise.

(3) A stepped HWP is of no value for reducing the $1/f$ excess noise considered in this paper, 
although a stepped HWP can help reduce other errors (e.g., from asymmetric beams) that are
important but not the topic of this study. 

(4) Finally, we close with some caveats concerning the applicability of these idealized results 
to future surveys. All the results above are rigorous and exact, but the rules of the game used 
here may not perfectly describe actual future surveys because of some of the assumptions made 
in the modelling. We have assumed a beam profile that is azimuthally symmetric about the beam 
center. We have also assumed that in the case in which the results of many detectors need to be 
combined in order to render the survey isotropic, all the beams have exactly the same profile. When 
each detector individually provides an isotropic survey, separate sky maps may be constructed 
for each detector and subsequently averaged over. However when this is not possible, additional 
errors are introduced due to the beam profile mismatch. Moreover frequency band mismatch 
introduces additional errors, and uncorrected errors from gain drifts have not been taken into 
account. Although we have not explored this question numerically, we do not believe that a 
mildly anisotropic survey will have errors significantly different from those calculated here 
for an isotropic survey. 
Certainly, at least within the rules of the game established in this paper, an isotropic survey that has
been rendered anisotropic by adding more scans will be more informative and will never have more
noise than the smaller isotropic subset of the data.
The main difference is that an anisotropic survey will be 
computationally much more demanding to analyze.

(5) In the case of a scanning pattern that is mildly anisotropic, the
isotropic approximation to the operator $A^TN^{-1}A$ could also be useful
as a preconditioner for solving the map making equation using the
conjugate gradient method.

(6) Overall, we observe that for temperature anisotropy measurements on 
the largest scales on the sky not to be compromised by 
$1/f$ noise (i.e., with a boost factor only slightly above 
unity), the beam must traverse most of the sky within a 
timescale of order ${\nu _{knee}}^{-1}$ or shorter. 
There is no alternative. This requires large scanning circles. 
For polarization, however, it is possible, at least within the 
framework of the idealizations assumed in this paper, 
even for small scanning circles, to reduce the $1/f$ noise boost 
on all angular scales by spinning the satellite or using a rotating 
half-wave plate when the angular velocity is of order $\nu _{knee}$ 
or larger. This difference can be understood from the absence of 
oscillations in the integrand of eqn.~(\ref{wTemp}) other than those 
oscillations arising from the Legendre polynomials---that is, from 
the mode functions themselves. For the polarization case, by 
contrast, the spinning of the satellite or the rotation of the HWP 
causes the integrand of eqn.~(\ref{wPol}) to oscillate, introducing 
cancellations in addition to those arising from the mode functions 
of the sky signal. The boost factors calculated in this paper should 
be understood as minimum values for the noise in the final maps to 
be expected from a real experiment. There is no possibility to do 
better. However further less idealized studies taking into account 
more sources of systematic error are needed before one can make 
reliable forecasts for realistic experiments. The map making equation 
may contain subtle cancellations that become spoiled by other systematic 
effects not included in this study.  

\vskip 4pt 

\noindent
{\bf Acknowledgements:} The author thanks 
Mark Ashdown,
Ken Ganga, Kavilan Moodley,
Guillaume Patachon, Michel Piat, 
Jonathan Sievers, and 
George Smoot, and especially Hirokazu Ishino 
for useful discussions and comments.
MB also thanks Kavilan Moodley for help with the figures.

\appendix 
\section{Explicit form of polarization mode functions}
\label{myAppendix}

We now give the explicit form of the eigenfunctions $E_{ab,j2}$ and $B_{ab,j2}.$ 
The $B_{ab,j2}$ eigenfunction is related to $E_{ab,j2}$ by a trivial $45^\circ $ 
rotation and has the same eigenvalue; therefore, we restrict ourselves to the 
explicit form of $E_{ab,j2},$ generated from $Y_{j2}(\hat \Omega )$
in the following way \cite{reionBump}: 
\begin{equation}
E_{ab,j2} =
\sqrt{\frac{2}{(l+2)(l+1)l(l-1)}}
\Bigl( \nabla _a \nabla _b -\frac{1}{2}\delta _{ab}\nabla ^2 \Bigr) 
Y_{j2}(\hat \Omega )
\label{eModeFromPotential}
\end{equation}
where $\nabla _a$ is the covariant derivative on the sphere 
with respect to the orthonormal basis labelled by $a,b,$ and
[DLMF, eqn.~(14.30.1)]
\begin{eqnarray}
&&Y_{j2}(\hat \Omega )=
\left( \frac{(l-m)!(2l+1)}{4\pi (l+m)!} \right) ^{1/2}
P_j^2(\cos \theta )~\exp (+2i\phi )\phantom{\int}
\end{eqnarray}
where $P_j^2(z)$ is the associated Legendre function of the first kind (or Ferrers function of the first kind). 
(For a compilation of properties of the associated Legendre 
functions, see Abramowitz and Stegun \cite{abramowitz}, Chapter 8 or DLMF, Chapter 14 \cite{dlmf}.)

The recurrence relation [eqn.~(14.10.05) of \cite{dlmf}]
\begin{equation}
(1-z^2) \frac{~d}{dz}P_l^m(z)= (l+m) P_{l-1}^m(z) -l P_l^m(z)
\end{equation}
is also useful.

In terms of the familiar basis 
$\mathbf{t}_{+}=
(\hat {\boldsymbol{\theta }} \otimes \hat {\boldsymbol{\theta }}
-\hat{\boldsymbol{\phi }}  \otimes \hat {\boldsymbol{\phi } })/\sqrt{2},$
and 
$\mathbf{t}_{\times }=
 (\hat {\boldsymbol{\theta }}\otimes \hat {\boldsymbol{\phi }} 
+ \hat {\boldsymbol{\phi }} \otimes  \hat {\boldsymbol{\theta }})/\sqrt{2},$
where
$(\hat {\boldsymbol{\theta }}, \hat {\boldsymbol{\phi }})=
\Bigl( (\partial /\partial \theta ), (1/\sin \theta ) (\partial /\partial \phi )\Bigr)$
serves as an orthonormal basis on the sphere, we have \cite{reionBump}
\begin{eqnarray}
\mathbf{E}_{jm}&=&
\sqrt{\frac{(2j+1)}{\pi }}
\sqrt{\frac{
(j-2)!(j-m)!
}{
(j+2)!(j+m)!
}}~
\exp [im\phi ]\cr 
&&
\qquad
\times \Biggl[
\left\{ 
~ (-)\left( \frac{j-m^2}{\sin ^2 \theta } +\frac{j(j-1)}{2}\right) P_j^m(\cos \theta )
+(j+m)~\frac{\cos \theta }{\sin ^2\theta }~P^m_{j-1}(\cos \theta) 
\right\} 
\mathbf{t}_{+} 
\cr
&&
\qquad 
\qquad 
\qquad 
\quad 
+i
\left\{ 
~ \frac{-m^2}{\sin ^2 \theta } \left[ \phantom{\bigg|} (j-1)\cos \theta ~
P_j^m(\cos \theta )
-(j-m)~P^m_{j-1}(\cos \theta) \right] 
\right\} 
\mathbf{t}_{\times }~
\Biggr] \cr 
&=&
Q_{jm}(\cos \theta )~\exp \bigl[ im\phi \bigr] ~
\mathbf{t}_{+}~
+U_{jm}(\cos \theta )~ \exp \bigl[ im\phi \bigr]~ \mathbf{t}_{\times }.\phantom{\int }
\end{eqnarray}
We have normalized so that
\begin{equation}
\sum _{a,b=1}^2 \int _0^\pi d\theta ~\sin \theta \int _0^{2\pi }d\phi ~
E^*_{ab,~jm}(\theta , \phi )~
E_{ab,~j'm'}(\theta , \phi ) = \delta _{jj'}~\delta _{mm'},
\end{equation}
which translates into the following normalization condition for the mode functions $Q_{jm}(z)$
and $U_{jm}(z)$:
\begin{equation}
(2\pi )\int _{-1}^{+1}dz~
\Bigl[
\left| Q_{jm}(z)\right| ^2
+\left| U_{jm}(z)\right| ^2
\Bigr] =1.
\end{equation}

We use the relation [DLMF, 14.3.4]
\begin{eqnarray}
P_j^m(x)=
(-1)^m\frac{
\Gamma (j+m+1)
}{
2^m
\Gamma (j-m+1)
}~ 
(1-x^2)^{m/2}~
\frac{1}{\Gamma (m+1)}~~
{}_2F_1\left( j+m+1, m-j; m+1; 
\frac{1}{2}
-
\frac{x}{2}\right)
\end{eqnarray}
to explore the behavior in the neighborhood of the north pole, obtaining the following
approximation 
\begin{eqnarray}
P_j^2(x)=\frac{1}{8} (j+2) (j+1) j (j-1) (1-x^2)
\end{eqnarray}
valid for $x\approx 1.$
It follows that
\begin{eqnarray}
\mathbf{E}_{j2}(\theta =0+, \phi )
=\frac{1}{4}\sqrt{\frac{(2j+1)}{\pi }}
(\mathbf{t}_{+}-i\!~\mathbf{t}_\times ) \exp [2i\phi ] .
\end{eqnarray}

\end{document}